\documentclass[twocolumn,aps,pra,superscriptaddress]{revtex4}
\usepackage[latin9]{inputenc}
\usepackage{color}
\usepackage{amsmath}
\usepackage{amssymb}
\usepackage{graphicx}
\usepackage[unicode=true,
 bookmarks=false,
 breaklinks=false,pdfborder={0 0 1},backref=section,colorlinks=true]
 {hyperref}
\hypersetup{
 linkcolor=blue,citecolor=blue,urlcolor=blue}
\usepackage{breakurl}

\makeatletter
\@ifundefined{textcolor}{}
{%
 \definecolor{BLACK}{gray}{0}
 \definecolor{WHITE}{gray}{1}
 \definecolor{RED}{rgb}{1,0,0}
 \definecolor{GREEN}{rgb}{0,1,0}
 \definecolor{BLUE}{rgb}{0,0,1}
 \definecolor{CYAN}{cmyk}{1,0,0,0}
 \definecolor{MAGENTA}{cmyk}{0,1,0,0}
 \definecolor{YELLOW}{cmyk}{0,0,1,0}
}

\usepackage{bm}
\usepackage{color}
\usepackage{braket}
\usepackage{subfigure}\usepackage{epsfig}\usepackage{amsfonts}\usepackage{mathrsfs}\usepackage[toc,page,title,titletoc,header]{appendix}

\@ifundefined{textcolor}{}{%
 \definecolor{BLACK}{gray}{0}
 \definecolor{WHITE}{gray}{1}
 \definecolor{RED}{rgb}{1,0,0}
 \definecolor{GREEN}{rgb}{0,1,0}
 \definecolor{BLUE}{rgb}{0,0,1}
 \definecolor{CYAN}{cmyk}{1,0,0,0}
 \definecolor{MAGENTA}{cmyk}{0,1,0,0}
 \definecolor{YELLOW}{cmyk}{0,0,1,0}
}

\makeatother

\begin{document}

\title{Two-body bound state of ultracold Fermi atoms with two-dimensional
spin-orbit coupling}

\author{Shu Yang}
\affiliation{Department of Physics, Renmin University of China, Beijing, 100872,
China}

\author{Fan Wu }
\email{wufan6366@mail.tsinghua.edu.cn}

\affiliation{Department of Physics, Tsinghua University, Beijing 100084, China}

\author{Wei Yi}
\email{wyiz@ustc.edu.cn}
\affiliation{CAS Key Laboratory of Quantum Information, University of Science and Technology of China, Hefei 230026, China}
\affiliation{CAS Center For Excellence in Quantum Information and Quantum Physics}

\author{Peng Zhang}
\email{pengzhang@ruc.edu.cn}
\affiliation{Department of Physics, Renmin University of China, Beijing, 100872,
China}
\affiliation{Beijing Computational Science Research Center, Beijing, 100084, China}
\affiliation{Beijing Key Laboratory of Opto-electronic Functional Materials \&
Micro-nano Devices, Renmin Univeristy of China, Beijing 100872, China}

\begin{abstract}
In a recent experiment, a two-dimensional spin-orbit coupling (SOC) was realized for fermions in the continuum [Nat. Phys. {\bf 12}, 540 (2016)], which represents an important step forward in the study of synthetic gauge field using cold atoms.
In the experiment, it was shown that a Raman-induced two-dimensional SOC exists in the dressed-state basis close to a Dirac point of the single-particle spectrum. By contrast, the short-range inter-atomic interactions of the system are typically expressed in the hyperfine-spin basis. The interplay between synthetic SOC and interactions can potentially lead to interesting few- and many-body phenomena but has so far eluded theoretical attention. Here we study in detail properties of two-body bound states of such a system. We find that, due to the competition between SOC and interaction, the stability region of the two-body bound state is in general reduced. Particularly, the threshold of the lowest two-body bound state is shifted to a positive, SOC-dependent scattering length.
Furthermore, the center-of-mass momentum of the lowest two-body bound state becomes nonzero, suggesting the emergence of Fulde-Ferrell pairing states in a many-body setting.
Our results reveal the critical difference between the experimentally realized two-dimensional SOC and the more symmetric Rashba or Dresselhaus SOCs in an interacting system, and paves the way for future characterizations of topological superfluid states in the experimentally relevant systems.
\end{abstract}

\maketitle

\section{Introduction}

Spin-orbit coupling (SOC) plays crucial roles in a wide range of physical contexts including atomic fine structures, high-$T_c$ superconductors and topological matter~\cite{rev1,rev2,rev3}. The implementation of synthetic SOCs in cold atomic systems thus offers exciting possibilities of quantum simulation using cold atoms~\cite{1D1,1D2,1D3,1D4,1D5,1D6,1D7,1D8,1D9,1D10,1D11,xiongjunscience,zhangjingNP2016,zhangjingPRL2016,zhangjingPRA2017,xiongjunPRL2018,xiongjunPRL2018b,xiongjunTheoryPRL,xiongjunTheoryPRLb,xiongjun3D,socreview1,socreview2,socreview4,socreview5,socreview6,yinlan1,yinlan2,2body1,2body2,2body3,2body4,2body5,2body6,2body7,2body8,2body9,2body10,2body11,2body12,2body13,2body14,2body15,2body16,2body17,2body18,2body19,2body20}. Specifically, the recent experimental realizations of two-dimensional (2D) SOC~\cite{xiongjunscience,zhangjingNP2016,zhangjingPRL2016,zhangjingPRA2017,xiongjunPRL2018,xiongjunPRL2018b}
opens up the avenue of simulating topological phenomena in higher dimensions. For the 2D SOC realized using the Raman lattice~\cite{xiongjunscience,xiongjunPRL2018,xiongjunPRL2018b,xiongjunTheoryPRL,xiongjunTheoryPRLb}, the non-trivial topology of single-particle band structures gives rise to dynamic topological phenomena in quench processes and can lead to topologically non-trivial phases when the lowest band is filled with fermions. In contrast, for fermions in the continuum under a Raman-induced 2D SOC~\cite{zhangjingNP2016,zhangjingPRL2016,zhangjingPRA2017}, a topological superfluid phase may be stabilized by introducing a pairing gap at the Fermi surface. However, in these latter systems, the 2D synthetic SOC emerges in the dressed-state basis, whereas the short-range $s$-wave interaction potentials are diagonal in the hyperfine-spin basis.
The inter-atomic interactions thus acquire a complicated form in the dressed-state basis, which makes it difficult to have a direct understanding of pairing physics therein.

In this work, we study in detail properties of two-body bound states in an ultracold Fermi gas of three-component fermionic atoms with the Raman-induced 2D SOC implemented in Ref.~\cite{zhangjingNP2016,zhangjingPRL2016,zhangjingPRA2017}. For simplicity, we assume the $s$-wave inter-atomic interaction be non-negligible only when the two atoms are in two specific hyperfine states, which is naturally the case when the system is tuned close to an $s$-wave Feshbach resonance between these two states.
We then exactly solve the two-body problem of this system for various scattering lengths $a$, center-of-mass (CoM) momenta, as well as with different frequencies and intensiteis of the Raman beams.
Our work is therefore a first step toward a systematic understanding of the effects of interaction in these systems.

We focus on the impact of the SOC on the thresholds of two-body bound states.
For our system, the two-body bound state of a given CoM momentum ${\bf K}$ appears only when $1/a$ is larger than a threshold value $C_{\rm th}({\bf K})$, i.e., $1/a>C_{\rm th}({\bf K})$. In the absence of SOCs, it is well-known that $C_{\rm th}=0$.
In the presence of an SOC, the threshold $C_{\rm th}$ would be shifted, which signals the impact of SOC on the stability of two-body bound states.  For systems with a highly symmetric synthetic SOC, such as a 2D Rashba- or Dresselhaus-type SOC or a three-dimensional isotropic SOC, it has been shown that~\cite{shenoyPRB2011,hanpuNLP2013} $C_{\rm th}=-\infty$, i.e., the two-body bound state can appear for an arbitrary scattering length. Thus, the stability region of the two-body bound state is significantly extended by symmetric SOCs. Nevertheless, for experimental systems with Raman-induced one-dimensional SOC in the continuum~\cite{1D1,1D2,1D3,1D4,1D5,1D6,1D7,1D8,1D9,1D10,1D11}, previous theoretical~\cite{peng2013,Melo} studies show that $C_{\rm th}>0$, i.e., the stability region of the two-body bound state is reduced by the SOC. This result is verified experimentally in Ref.~\cite{spielmanPRL2013}.
All these studies show that different types of SOC can induce qualitatively different
modifications of $C_{\rm th}$. Therefore, it is necessary to investigate the influence of the Raman-induced 2D SOC on the threshold $C_{\rm th}$ of the two-body bound state. In this work, we calculate $C_{\rm th}$ for various cases under the experimentally implemented 2D SOC, and show that in each case $C_{\rm th}$
is always shifted to a {\it positive} value which depends on the CoM momentum, i.e., $C_{\rm th}({\bf K})>0$. This result shows that the stability region of the two-body bound state is reduced by the Raman-induced 2D SOC, similar to that under a Raman-induced one-dimensional SOC.

In addition, we investigate properties of the ``ground" two-body bound state, which has the lowest energy under fixed scattering length and Raman-beam-parameters.
We show that the binding energy of the ground bound state is smaller than that without Raman beams. Namely, the Raman-induced SOC makes the ground two-body bound state shallower.
Furthermore, the CoM momentum of this ground two-body bound state is nonzero and typically lies in the plane of the 2D SOC. One would therefore expect the emergence of Fulde-Ferrell pairing state in a many-body system, which would compete with the normal state.

Our results reveal that, whereas the Raman-induced 2D SOC can be symmetric in the dressed-state basis on the single-particle level, inter-atomic interactions break both the rotational and inversion symmetries, giving rise to less stable two-body bound states with finite CoM momenta. These phenomena can be understood by projecting inter-atomic interactions into the dress-state basis, where scattering states in the dressed states are momentum-dependent. Our work reveals the nontrivial interplay of SOC and interaction in an experimentally relevant system, and provides the necessary basis for future studies of the system on the many-body level.

\begin{figure}[t]
\begin{centering}
\includegraphics[width=8cm]{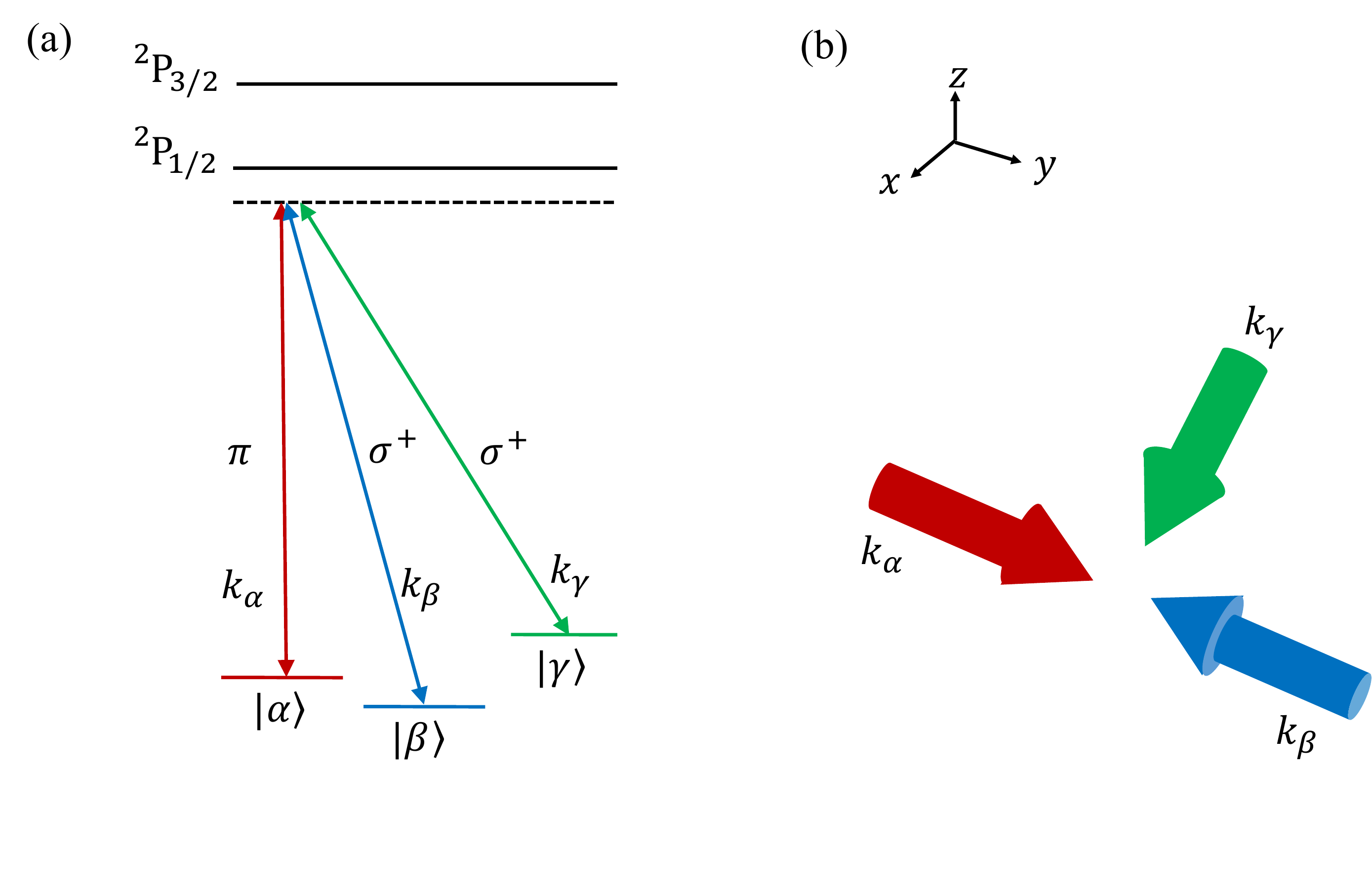}
\par\end{centering}
\caption{(color online) \textbf{(a)}: The schematic diagram of the Raman coupling
between the hyperfine states $\alpha,$ $\beta$ and $\gamma$.
In our system each atom has three internal states $\alpha$, $\beta$ and $\gamma$, and the two-body interaction appears when one atom is in state $\alpha$ and another atom is in state $\beta$.
 \textbf{(b)}:
The directions of the Raman laser beams. \label{scheme}}
\end{figure}

Our paper is organized as follows. In Sec.~\ref{sec:2}, we review the coupling scheme for the Raman-induced two-dimensional SOC and present our theoretical approach from which two-body bound-state energies are calculated.
 We investigate the effects of Raman-induced SOC on the threshold and stability region of the two-body bound state in Sec.~\ref{sec:3}. In Sec.~\ref{sec:4} we study the lowest two-body bound state. Finally, we summarize in Sec.~\ref{sec:5}. Some details of our calculation and analysis are given in the appendices.

\section{Calculation of two-body bound state}\label{sec:2}

We consider two identical ultracold Fermi atoms $1$ and $2$. As discussed in Ref.~\cite{zhangjingNP2016,zhangjingPRL2016,zhangjingPRA2017} and
shown in Fig.~\ref{scheme}, the ground
hyperfine states $\alpha,$ $\beta$ and $\gamma$ of each atom are coupled to
electronic-orbital excited manifolds ($^{2}{\rm P}_{1/2}$
and $^{2}{\rm P}_{3/2}$ manifolds) via far-off-resonant Raman
laser beams propagating along different directions and with wave vectors
\begin{eqnarray}
{\bf k}_{\alpha} = k_{r}{\bf e}_{y},\
{\bf k}_{\beta} = -k_{r}{\bf e}_{y},\
{\bf k}_{\gamma} = k_{r}{\bf e}_{x},\label{k3}
\end{eqnarray}
respectively. Here ${\bf e}_{j}$ ($j=x,y,z$) is the unit vector
in the $j$-direction. Notice that in the experiment, the norms of wave vectors of the three Raman beams are approximately same.

The Raman lasers couple the three hyperfine (internal) states in a pair-wise fashion. As a result, in the rotated frame, the free Hamiltonian $H^{(i)}$ for atom $i$
($i=1,2$) can be expressed as a function of its momentum ${\bf p}^{(i)}$
\begin{eqnarray}
H^{(i)}=H_{{\rm 1b}}({\bf p}^{(i)}) & \equiv & \sum_{\xi=\alpha,\beta,\gamma}\left[\frac{\left({\bf p}^{(i)}-{\bf k}_{\xi}\right)^{2}}{2m}+\delta_{\xi}\right]|\xi\rangle_{i}\langle\xi|\nonumber \\
 &  & -\sum_{\xi,\eta=\alpha,\beta,\gamma}\frac{\Omega_{\xi\eta}}{2}|\xi\rangle_{i}\langle\eta|,
 \label{hi}
\end{eqnarray}
where $m$ is the single-atom mass, $|\xi\rangle_{i}$
($\xi=\alpha,\beta,\gamma$) is the internal state of atom $i$, $\delta_{\xi}$
($\xi=\alpha,\beta,\gamma$) is the effective energy of the state $|\xi\rangle_{i}$, determined by the detuning of the laser beams.
$\Omega_{\xi\eta}$ ($\xi,\eta=\alpha,\beta,\gamma$) is the effective
Rabi frequency for the Raman transition between states $|\xi\rangle_{i}$
and $|\eta\rangle_{i}$, which satisfies $\Omega_{\xi\eta}=\Omega_{\eta\xi}^{\ast}$.
Eq.~(\ref{hi}) shows that the atomic momentum in the $x-y$ plane
is coupled to the internal states via terms proportional to ${\bf k}_{\alpha,\beta,\gamma}$, which amounts to
the Raman-induced 2D synthetic SOC.


To investigate two-body bound states under this configuration, we write the total Hamiltonian as
\begin{equation}
H=H_{F}+U,\label{h}
\end{equation}
with $H_{F}=H_{{\rm 1b}}({\bf p}^{(1)})+H_{{\rm 1b}}({\bf p}^{(2)})$
being the two-body free Hamiltonian and $U$ being the inter-atomic
interaction. In our system, the CoM momentum ${\bf K}={\bf p}^{(1)}+{\bf p}^{(2)}$
is conserved, and thus can be treated as a classical parameter (c-number).
Accordingly, the two-body free-Hamiltonian can be expressed as
\begin{equation}
H_{F}=H_{{\rm 1b}}\left(\frac{{\bf K}}{2}+{\bf p}\right)+H_{{\rm 1b}}\left(\frac{{\bf K}}{2}-{\bf p}\right),\label{hf}
\end{equation}
where ${\bf p}=({\bf p}_{1}-{\bf p}_{2})/2$ is the relative-momentum
operator of the two atoms.

When ${\bf K}$ is fixed, we only
need to study the quantum state of the two-atom relative spatial motion,
as well as the internal states of the atoms. Thus, the Hilbert
space ${\cal H}$ of our system is given by ${\cal H}={\cal H}_{r}\otimes{\cal H}_{s1}\otimes{\cal H}_{s2}$,
with ${\cal H}_{r}$ being the Hilbert space for the two-atom spatial
relative motion, and ${\cal H}_{si}$ ($i=1,2$) being the one for
internal states of atom $i$. Similar to our previous work~\cite{peng2013},
we use the symbol $|\rangle\!\rangle$ to denote states in
${\cal H}$, $|)$ for states in ${\cal H}_{{\rm motion}}$, $|\rangle_{i}$
($i=1,2$) for states in ${\cal H}_{si}$ and $|\rangle$ for states
in ${\cal H}_{s1}\otimes{\cal H}_{s2}$.


Furthermore, we assume that when one atom is in the internal state $\alpha$
and the other in $\beta$, the inter-atomic interaction is strong
and can be described by the $s$-wave scattering length $a$; whereas for all other
cases the interactions are negligible. One can experimentally realize
such a configuration by tuning the magnetic field close to the Feshbach resonance
of the hyperfine-spin channel corresponding to the singlet state
\begin{equation}
|S\rangle=\frac{1}{\sqrt{2}}\left(|\alpha\rangle_{1}|\beta\rangle_{2}-|\beta\rangle_{1}|\alpha\rangle_{2}\right),\label{singlet}
\end{equation}
where $a$ becomes magnetic-field-dependent. As we have proved
in Ref.~\cite{peng2013}, regardless of the presence of SOC, the
inter-atomic interaction can always be described by the widely-used
renormalized contact interaction, with
\begin{equation}
U=\frac{U_{0}}{(2\pi)^{3}}|S\rangle\langle S|\otimes\int_{k,k'<k_{c}}|{\bf k})({\bf k}'|d{\bf k}d{\bf k}'.\label{U}
\end{equation}
Here $|{\bf k})$ is the eigen-state of the relative-momentum operator
${\bf p}$, and the momentum cutoff $k_{c}$ satisfies the renormalization
relation ($\hbar=1$)~\cite{rmpfermi,uu,peng2013}:
\begin{equation}
\frac{m}{4\pi a}=\frac{1}{U_{0}}+\frac{m}{(2\pi)^{3}}\int_{k''<k_{c}}\frac{1}{k^{\prime\prime2}}d{\bf k}''.\label{renor}
\end{equation}

Now we calculate the energy  ${\cal E}_{b}$ of the two-body bound state $|\Psi_{b}\rangle\rangle$. With straightforward calculations (Appendix A) based on the Schr$\ddot{{\rm o}}$dinger equation
\begin{equation}
H|\Psi_{b}\rangle\rangle={\cal E}_{b}|\Psi_{b}\rangle\rangle,\label{se}
\end{equation}
we find that
\begin{equation}
\frac{1}{(2\pi)^{3}}\int d{\bf k}J[{\cal E}_{b},{\bf k};{\bf K}]=\frac{1}{4\pi a},\label{equa}
\end{equation}
where the function $J[{\cal E}_{b},{\bf k};{\bf K}]$ is defined as
\begin{eqnarray}
J[{\cal E}_{b},{\bf k};{\bf K}] & = & \sum_{\Lambda=1}^{9}\left[\frac{\left|\langle\Lambda,{\bf k},{\bf K}|S\rangle\right|^{2}}{m\left({\cal E}_{b}-E_{\Lambda,{\bf k},{\bf K}}\right)}+\frac{\left|\langle\Lambda,{\bf k},{\bf K}|S\rangle\right|^{2}}{k^{2}}\right].\nonumber \\
\label{jjj}
\end{eqnarray}
Here the two-body hyperfine spin state $|\Lambda,{\bf k},{\bf K}\rangle$
($\Lambda=1,2,...,9$) is the $\Lambda$-th eigen state of the operator
$h({\bf k},{\bf K})\equiv H_{{\rm 1b}}\left(\frac{{\bf K}}{2}+{\bf k}\right)+H_{{\rm 1b}}\left(\frac{{\bf K}}{2}-{\bf k}\right)$
and $E_{\Lambda,{\bf k},{\bf K}}$ is the corresponding eigen-energy.
Notice that in the definition of  $h({\bf k},{\bf K})$, both ${\bf K}$ and ${\bf k}$ are c-numbers, and thus
$h({\bf k},{\bf K})$ is an operator in the nine-dimensional Hilbert
space ${\cal H}_{s1}\otimes{\cal H}_{s2}$ \cite{hh}.

In addition, it is clear
that the two-body bound-state energy ${\cal E}_{b}$ should also satisfy
\begin{equation}
{\cal E}_{b}\leq E_{{\rm th}}({\bf K}),\label{con}
\end{equation}
where the threshold energy $E_{{\rm th}}({\bf K})$ is defined as
the lowest eigen-energy of the two-body free Hamiltonain $H_{F}$
for a fixed ${\bf K}$, i.e.,
\begin{eqnarray}
E_{{\rm th}}({\bf K})={\rm Min}[E_{\Lambda,{\bf k},{\bf K}}].
\end{eqnarray}

We numerically solve Eq.~(\ref{jjj}) under the condition (\ref{con}),
and obtain the bound-state energy ${\cal E}_{b}$ for each case with
given values of scattering length $a$, Rabi frequencies $\Omega_{\xi\eta}$
($\xi,\eta=\alpha,\beta,\gamma$), effective energies $\delta_{\xi}$
($\xi=\alpha,\beta,\gamma$) and CoM momentum ${\bf K}$.

\section{Threshold of two-body bound state}\label{sec:3}

In many systems, thresholds exist for two-body bound states where bound states only appear beyond them.
In our system, two-body bound states only appear when $1/a$ is larger than a threshold which we denote as $C_{\rm th}$, i.e.,
\begin{equation}
\frac{1}{a}\geq C_{\rm th}.\label{cck}
\end{equation}
In the absence of SOCs, it is well-known that $C_{\rm th}=0$ for $s$-wave interactions in three dimensions. In
the presence of the SOC, the location of the threshold $C_{\rm th}$ depends on the form of SOC, the dimensionality of the system, as well as the CoM momentum of the bound state. Specifically, for the Raman-induced 2D SOC considered here, the threshold depends on the $x$- and
$y$-components of the CoM momentum ${\bf K}$, i.e., $C_{\rm th}=C_{\rm th}({\bf K})=C_{\rm th}(K_{x},K_{y})$.

\begin{figure}[t]
\begin{centering}
\includegraphics[width=4cm]{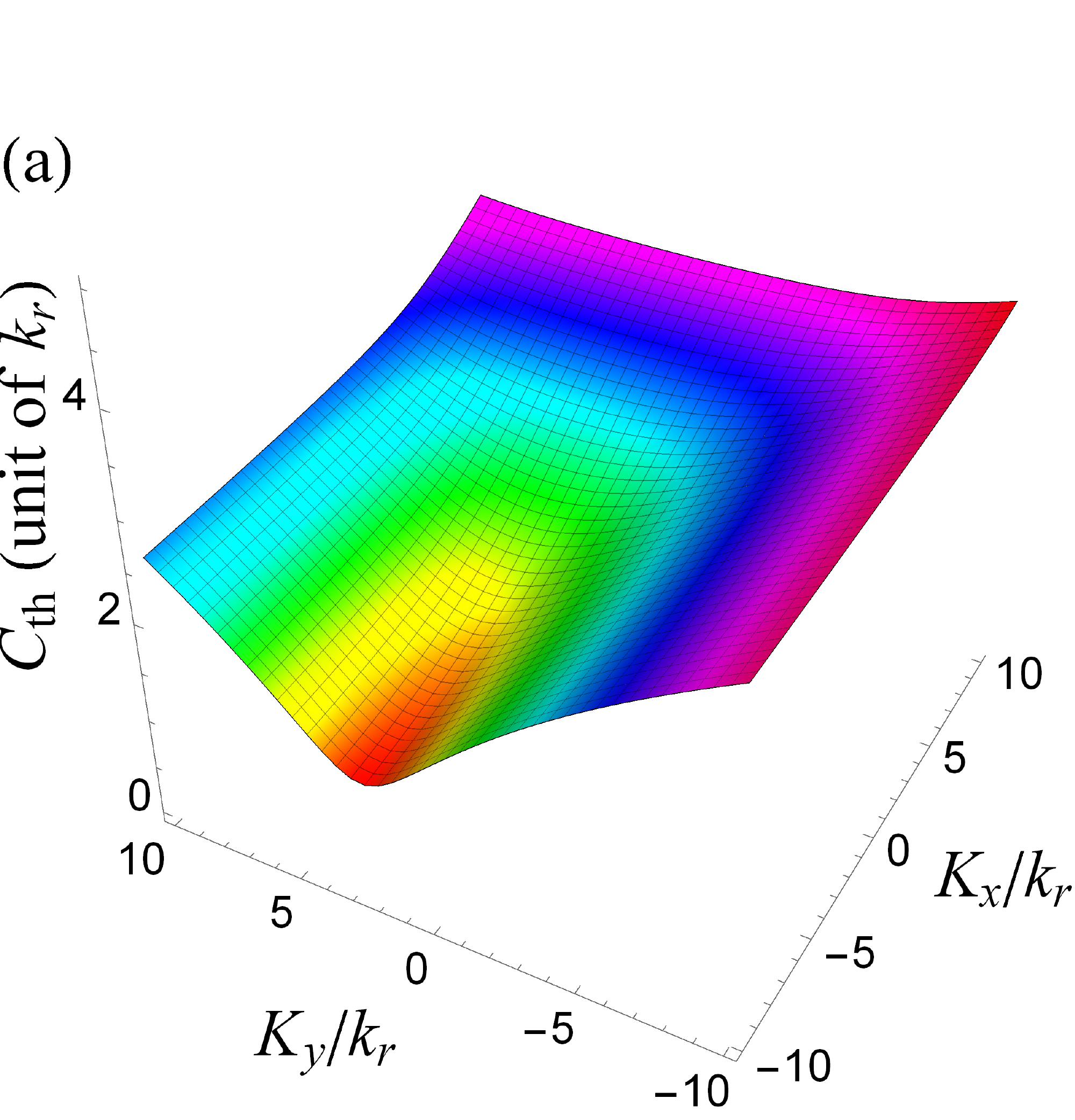}
\includegraphics[width=4cm]{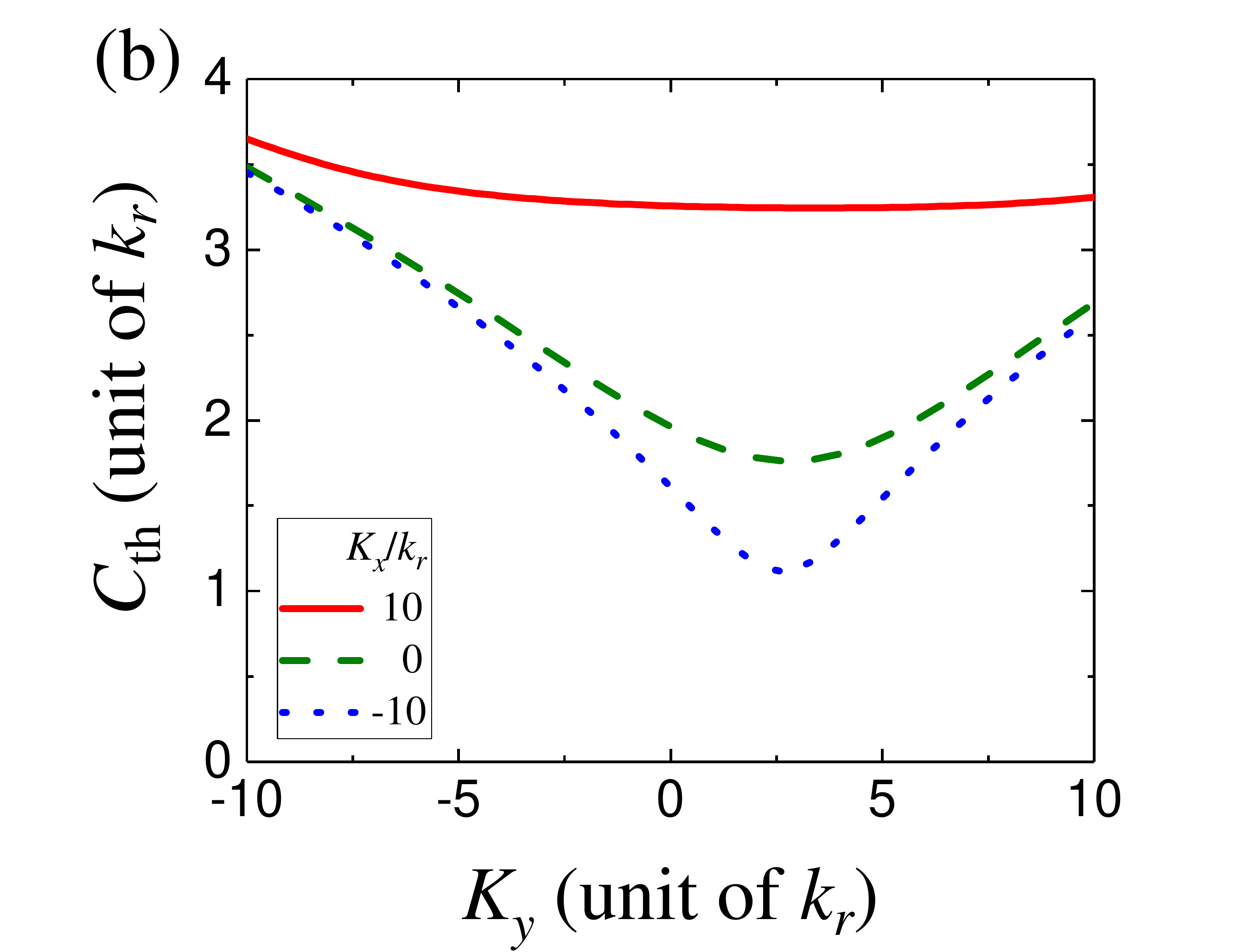}
\includegraphics[width=4cm]{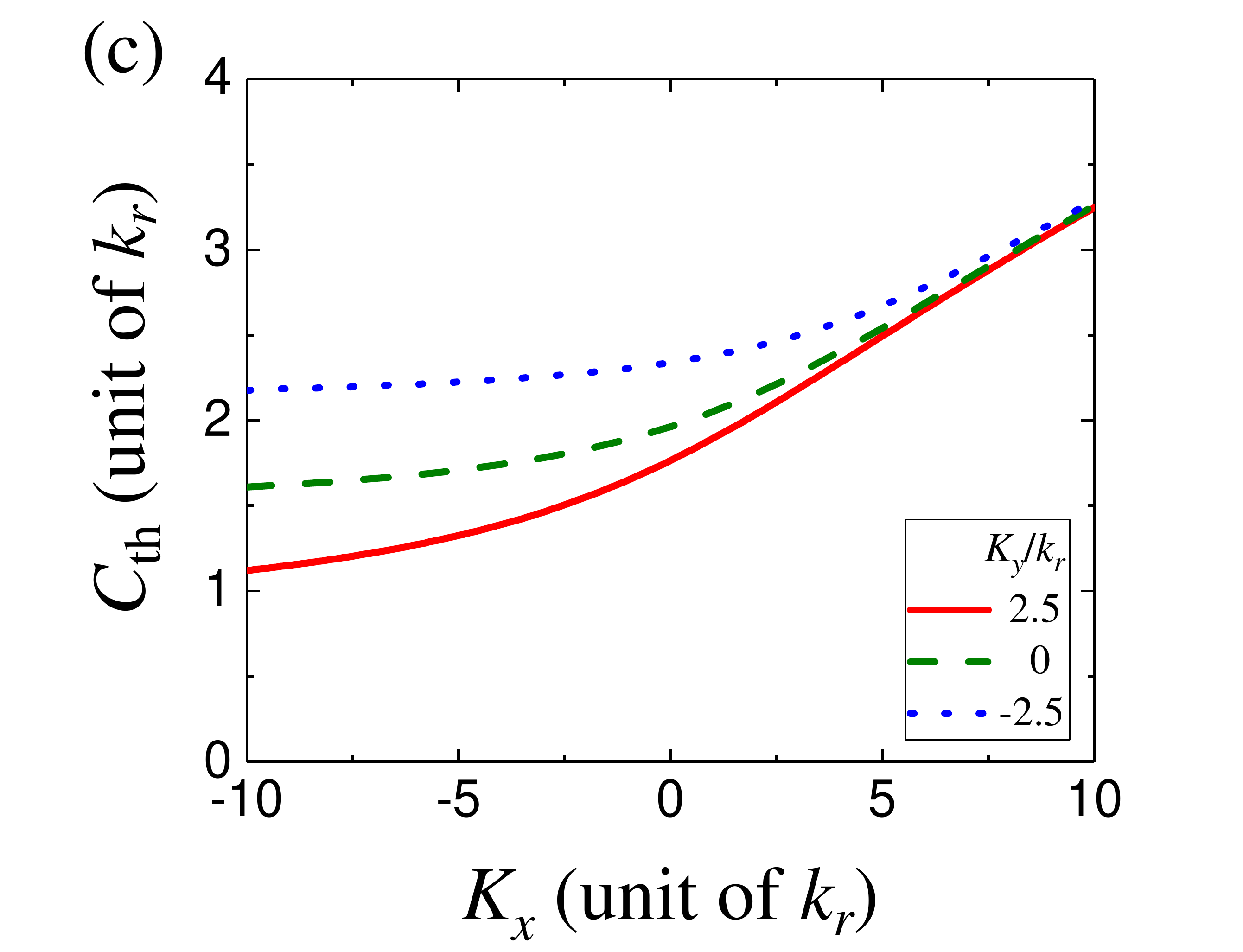}
\includegraphics[width=4cm]{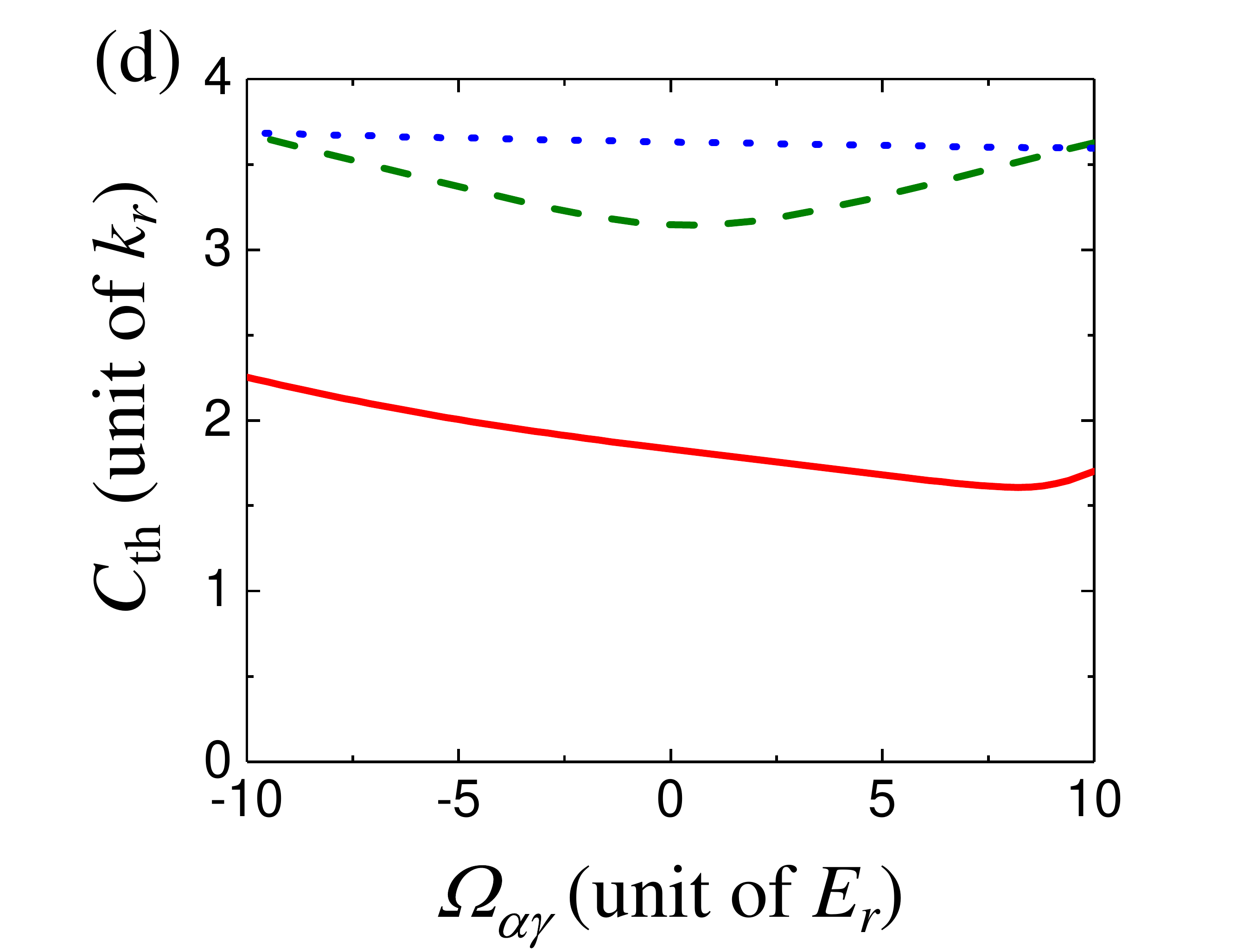}
\par\end{centering}
\caption{(color online) The two-body bound-state threshold $C_{\rm th}$ for systems under the scheme of Ref.~\cite{zhangjingNP2016}, with typical experimental parameters: $\Omega_{\alpha\beta}=3.58E_{r}$, $\Omega_{\alpha\gamma}=-3.94E_{r}$,
$\Omega_{\beta\gamma}=-4.66E_{r}$, $\delta_{\alpha}=0$,
$\delta_{\beta}=-5.14E_{r}$ and $\delta_{\gamma}=-3.23E_{r}$.
\textbf{(a)}:
$C_{\rm th}$ as a function of CoM momentum $(K_{x},K_{y})$.
\textbf{(b)}:
$C_{\rm th}$ as a function $K_{y}$, for $K_x=10k_r$ (red solid line), $K_x=0k_r$ (green dashed line), $K_x=-10k_r$ (blue dotted line).
\textbf{(c)}:
$C_{\rm th}$ as a function $K_{x}$, for $K_y=2.5k_r$ (red solid line), $K_y=0k_r$ (gren dashed line), $K_y=-2.5k_r$ (blue dotted line).
\textbf{(d)}: The threshold  $C_{\rm th}(K_x,K_y)$
as a function of the effective Rabi frequency $\Omega_{\alpha\gamma}$,
for cases with CoM momentum $(K_x,K_y)=(0,0)$ (red solid line), $(K_x,K_y)=(10k_r,10k_r)$ (green dashed line), and $(K_x,K_y)=(10k_r,-10k_r)$ (blue dotted line).
\label{ck}}
\end{figure}

\begin{figure}[t]
\begin{centering}
\includegraphics[width=4cm]{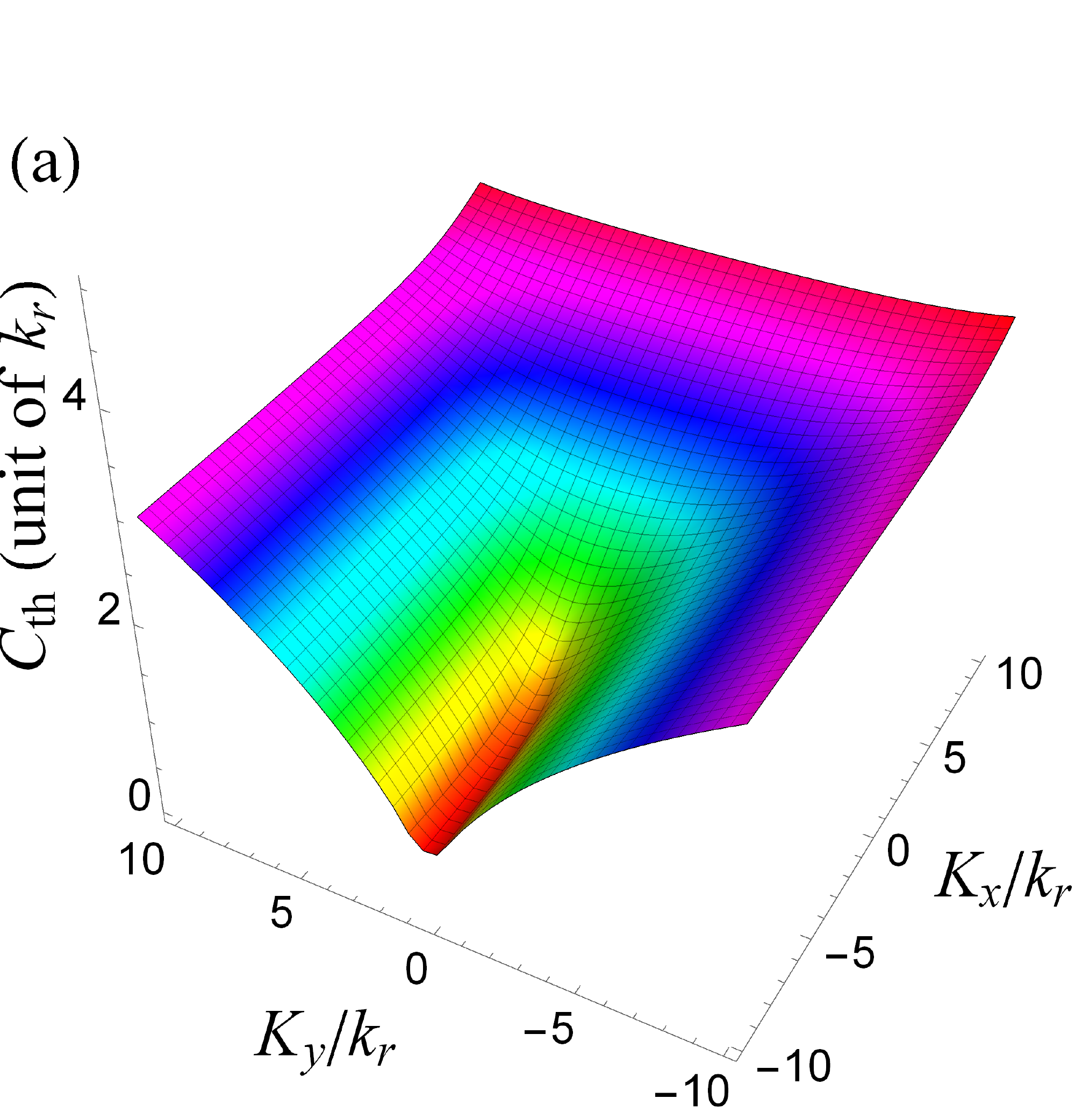}
\includegraphics[width=4cm]{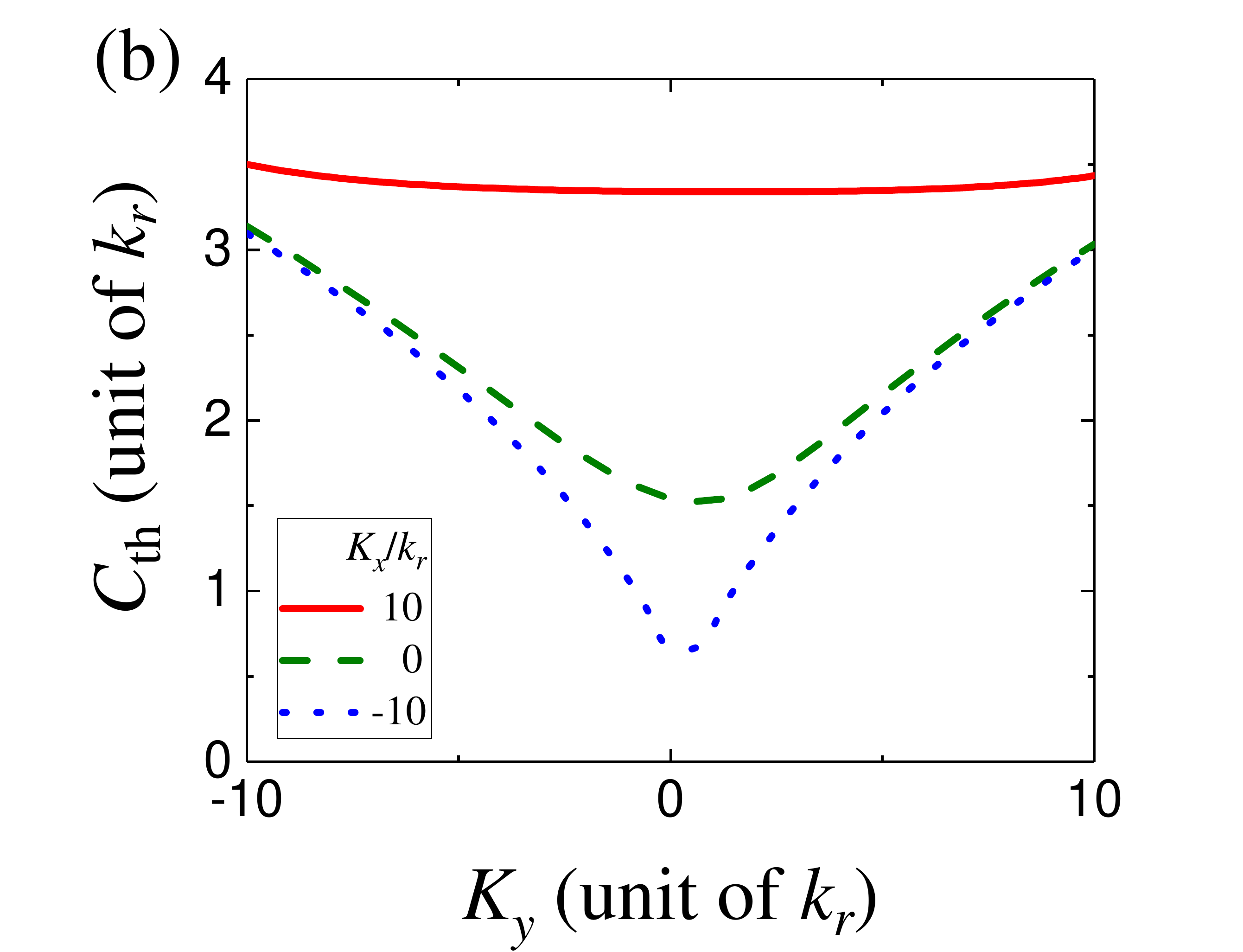}
\includegraphics[width=4cm]{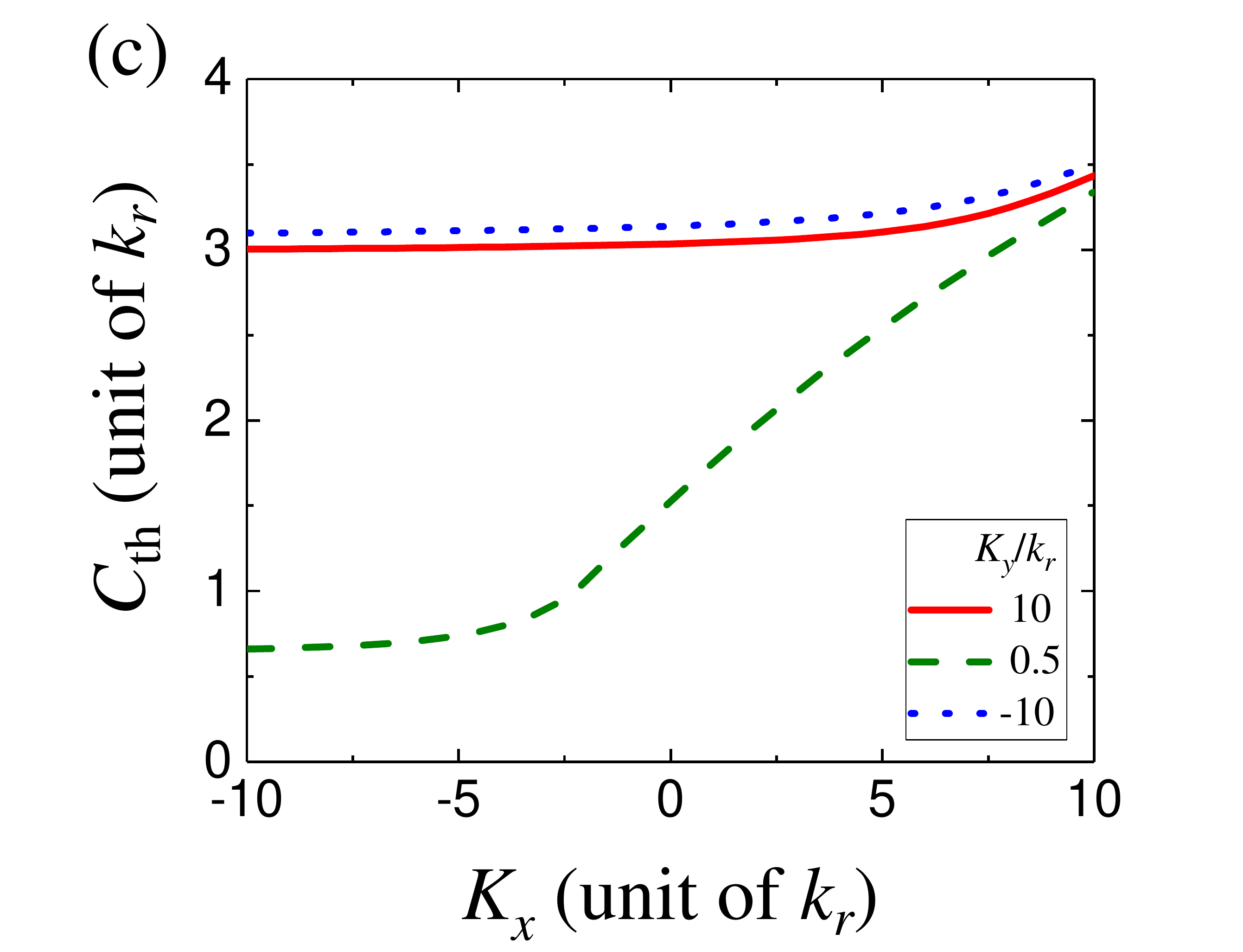}
\includegraphics[width=4cm]{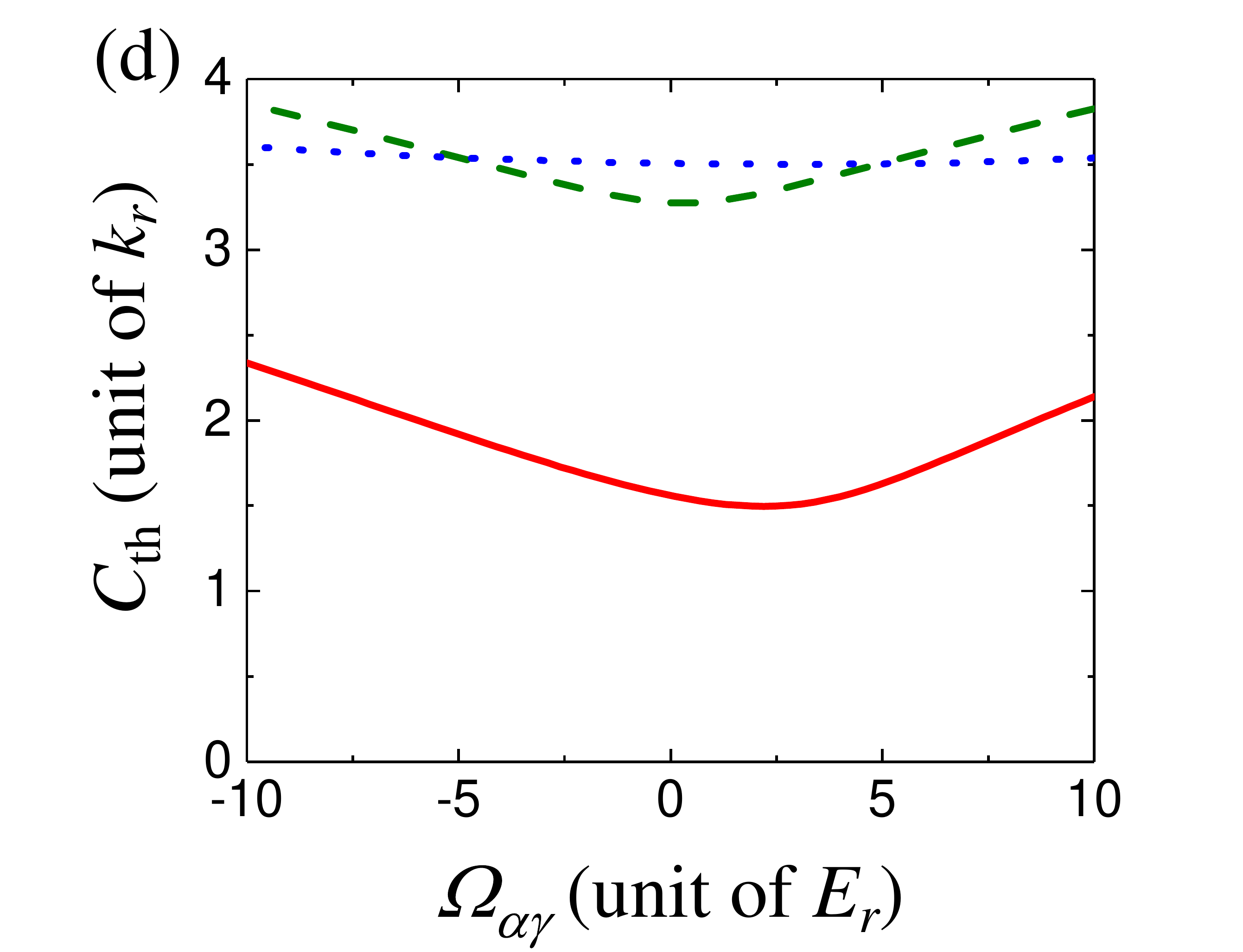}
\par\end{centering}
\caption{
(color online) The two-body bound-state threshold  $C_{\rm th}$ for systems under the scheme of
Ref.~\cite{zhangjingPRL2016}, with typical experimental parameters: $\Omega_{\alpha\beta}=-2.49(1+i) E_{r}$, $\Omega_{\alpha\gamma}=3.86E_r$,
$\Omega_{\beta\gamma}=4.58E_{r}$, $\delta_{\alpha}=0$,
$\delta_{\beta}=-0.5E_{r}$ and $\delta_{\gamma}=-1.8E_{r}$.
\textbf{(a)}:
$C_{\rm th}$ as a function of CoM momentum $(K_{x},K_{y})$.
\textbf{(b)}:
$C_{\rm th}$ as a function $K_{y}$, for $K_x=10k_r$ (red solid line), $K_x=0k_r$ (green dashed line), $K_x=-10k_r$ (blue dotted line).
\textbf{(c)}:
$C_{\rm th}$ as a function $K_{x}$, for $K_y=10k_r$ (red solid line), $K_y=0.5k_r$ (green dashed line), $K_y=-10k_r$ (blue dotted line).
\textbf{(d)}: The threshold $C_{\rm th}$
as a function of the effective Rabi frequency $\Omega_{\alpha\beta}$,
for cases with CoM momentum $(K_x,K_y)=(0,0)$ (red solid line), $(K_x,K_y)=(10k_r,10k_r)$ (green dashed line), and $(K_x,K_y)=(10k_r,-10k_r)$ (blue dotted line).
\label{ckc}}
\end{figure}

We numerically calculate $C_{\rm th}$ via the approach shown in Sec.~II and illustrate our results
for typical experimental parameters with real~\cite{zhangjingNP2016} and complex~\cite{zhangjingPRL2016} effective Rabi frequencies, respectively.
In Fig.~\ref{ck}(a-c) and Fig.~\ref{ckc}(a-c), we show $C_{\rm th}$ as a function of $(K_{x},K_{y})$, while in Fig.~\ref{ck}(d) and Fig.~\ref{ckc}(d), $C_{\rm th}$ is plotted as a function of $\Omega_{\alpha\gamma}$.
In addition, we systematically characterize the behavior of $C_{\rm th}$ with varying Rabi frequencies and detunings, and show the results in Appendix B.
All these results clearly show that, in the presence of SOC, we always have
\begin{equation}
C_{\rm th}(K_{x},K_{y})>0,\label{cp}
\end{equation}
which suggests that, on the ``$1/a$-axis\char`\"{}, the parameter regime for the existence of two-body bound states is reduced by
the Raman-induced SOC. In other words, the two-body bound state becomes
more difficult to form under the Raman-induced 2D SOC.

The result above is similar to those under Raman-induced one-dimensional SOCs~\cite{peng2013,Melo,spielmanPRL2013}.
In comparison, previous studies~\cite{shenoyPRB2011} on three-dimensional Fermi gases under the 2D Rashba- and Dresselhaus-type SOCs show that for such systems $C_{\rm th}=-\infty$, i.e., the two-body bound state can always be formed at an arbitrary scattering length. The stability region for the two-body bound state is therefore significantly broadened by the Rashba- or Dresselhaus-type SOCs. Thus, our result for the Raman-induced 2D SOC is in sharp contrast with these results, and highlight the critical difference between the Raman-induced 2D SOC and the more symmetric Rashba or Dresselhaus SOCs in interacting systems.

Furthermore, as illustrated in Fig.~\ref{ck}(a, b) and Fig.~\ref{ckc}(a, b), the bound-state threshold $C_{\rm th}$ increases with the CoM momentum $|K_y|$, i.e., the two-body bound state is more difficult to form when $|K_y|$ is large.
This can be understood as the following. In the two-body free Hamiltonian $H_{F}$, the component
$K_{y}$ contributes a term
\begin{eqnarray}
W\equiv\frac{K_{y}k_{r}}{2m}\sum_{j=1,2}\left(|\beta\rangle_{j}\langle\beta|-|\alpha\rangle_{j}\langle\alpha|\right),\label{ww}
\end{eqnarray}
which is proportional to the SOC intensity $k_r$. Under the influence of this term and for a large enough $|K_{y}|$, the threshold
energy $E_{{\rm th}}({\bf K})$ would decrease with increasing $|K_{y}|$.
As a result, the stability region of the bound state ${\cal E}_b$, shown in Eq.~(\ref{con}), would be reduced by the increase of $|K_{y}|$, rendering the two-body bound state more difficult to form.
For the similar reason, the
critical value $C_{\rm th}$ also becomes very large when $K_{x}$ takes a {\it positive} large value; and tends to a constant when $K_{x}$ takes a {\it negative} large value, as shown in Fig.~\ref{ck}(a)(c) and Fig.~\ref{ckc}(a)(c). In Appendix C, we show a more detailed explanation
for this physical picture.



\begin{figure}[t]
\begin{centering}
\includegraphics[width=4cm]{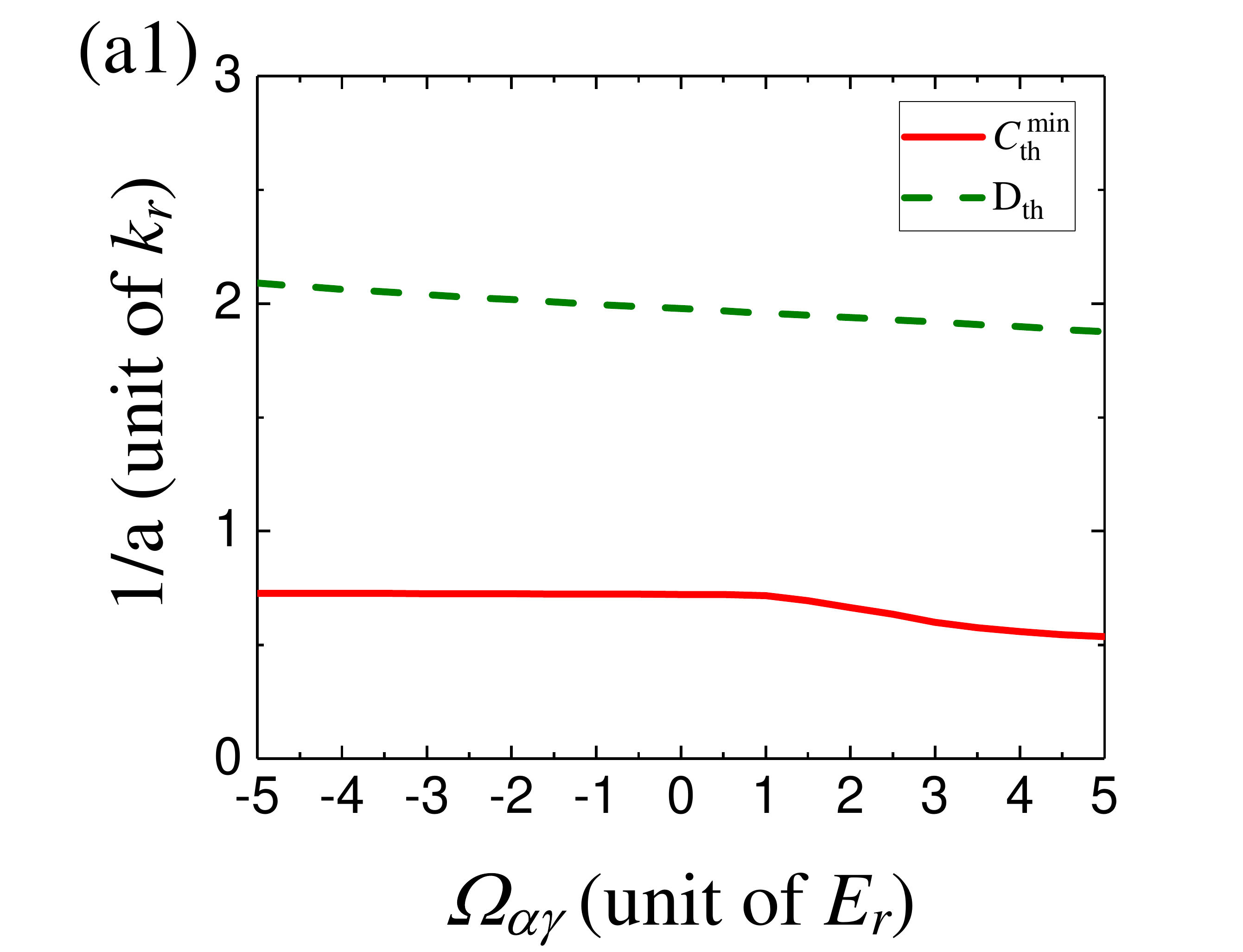}
\includegraphics[width=4cm]{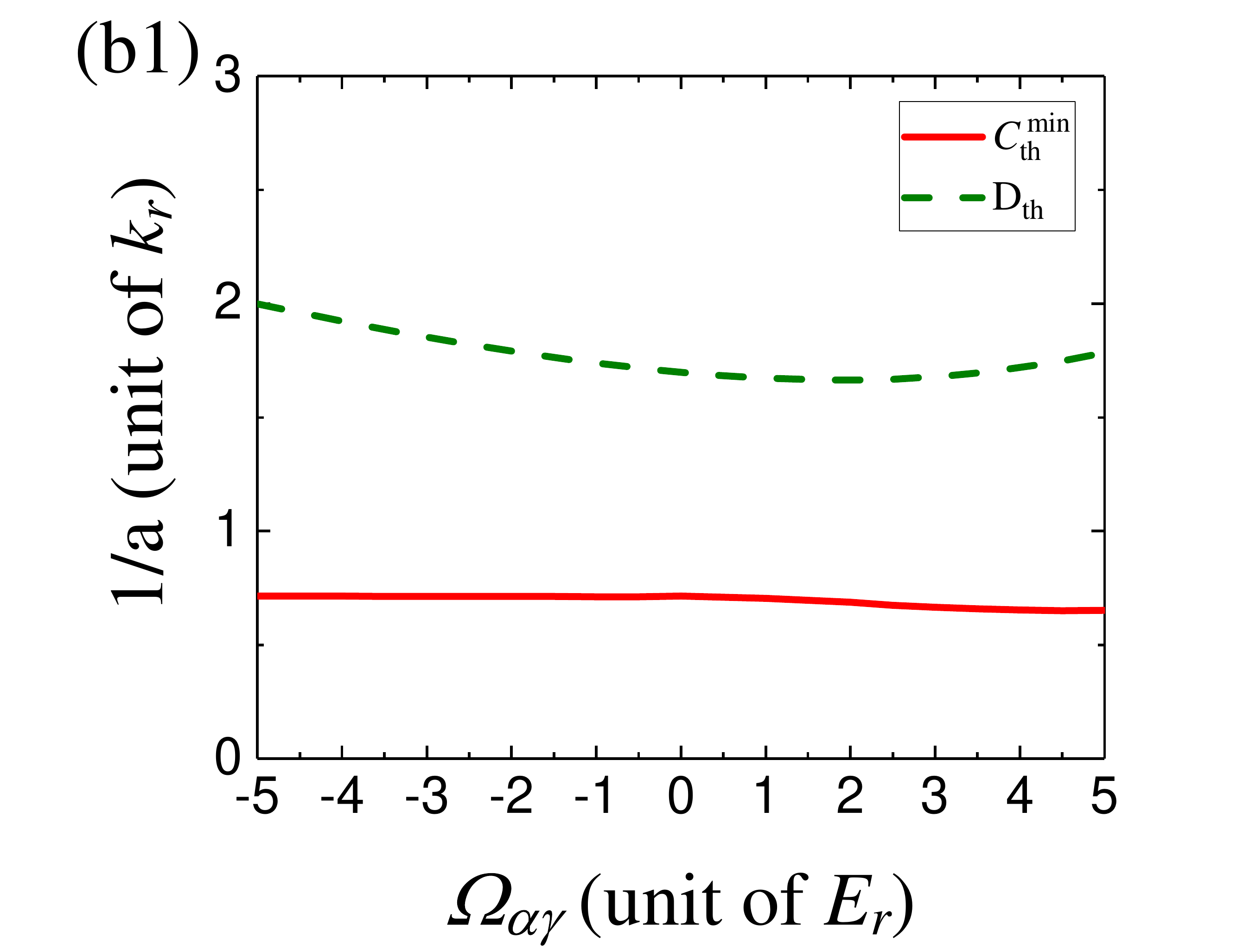}
\includegraphics[width=4cm]{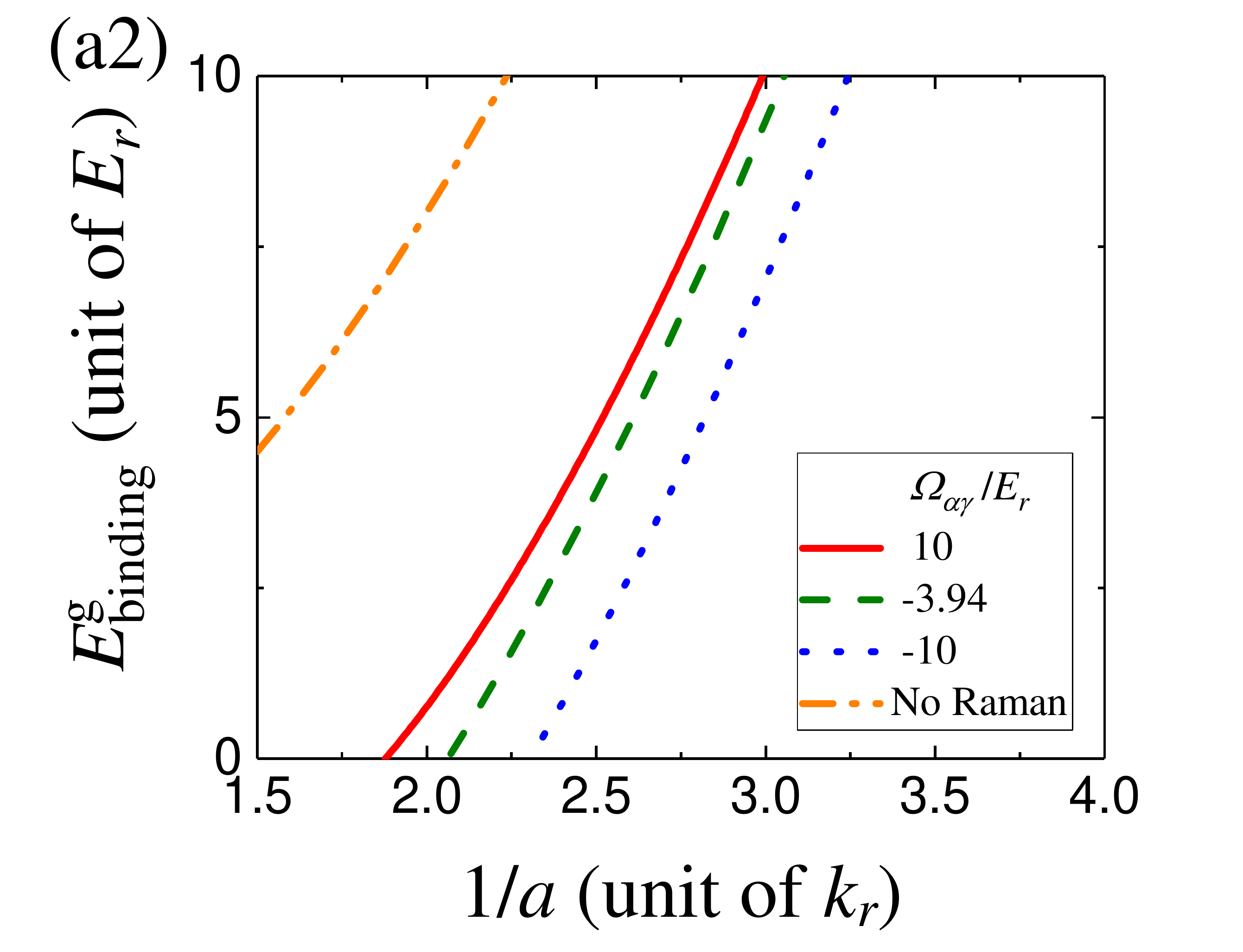}
\includegraphics[width=4cm]{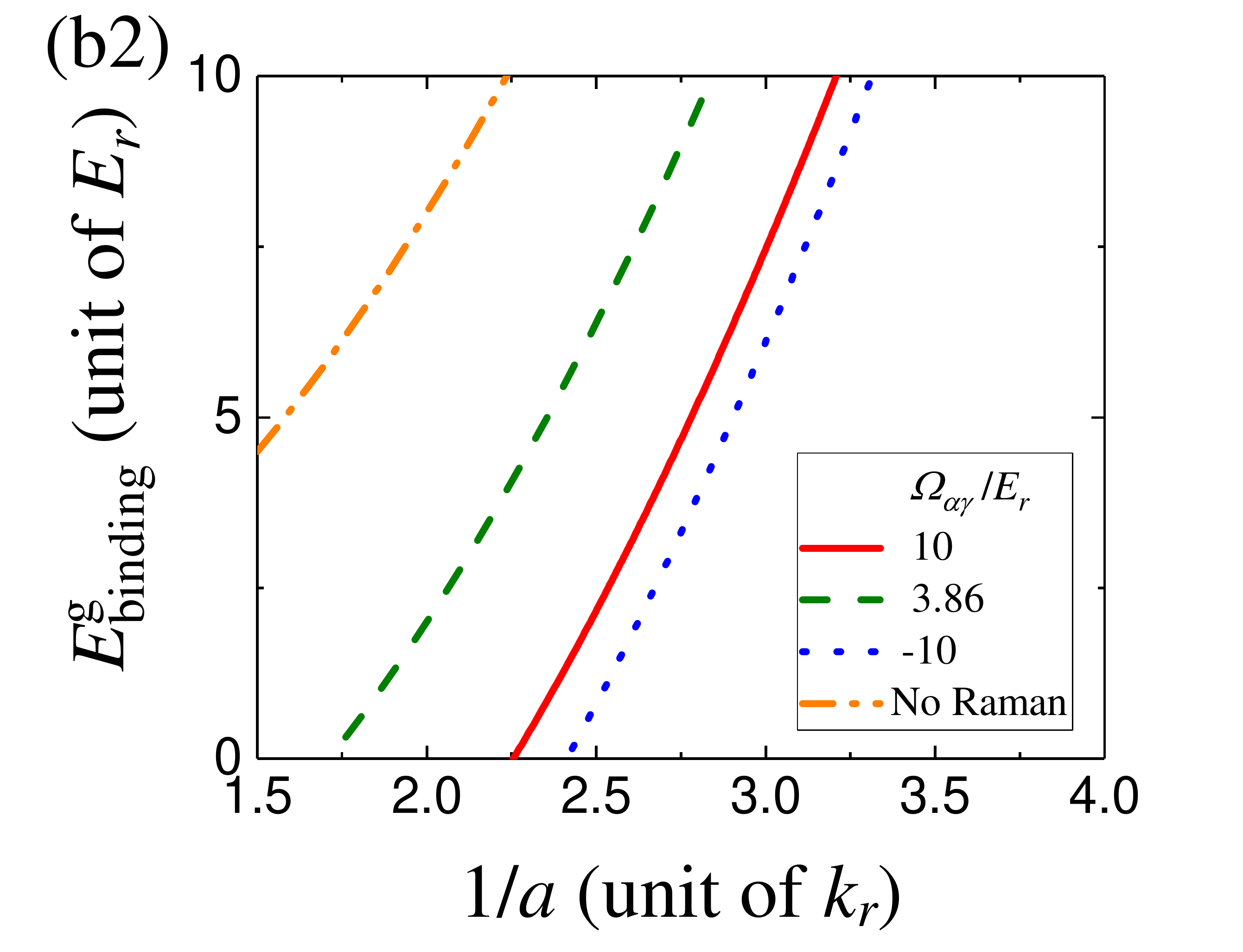}
\includegraphics[width=4cm]{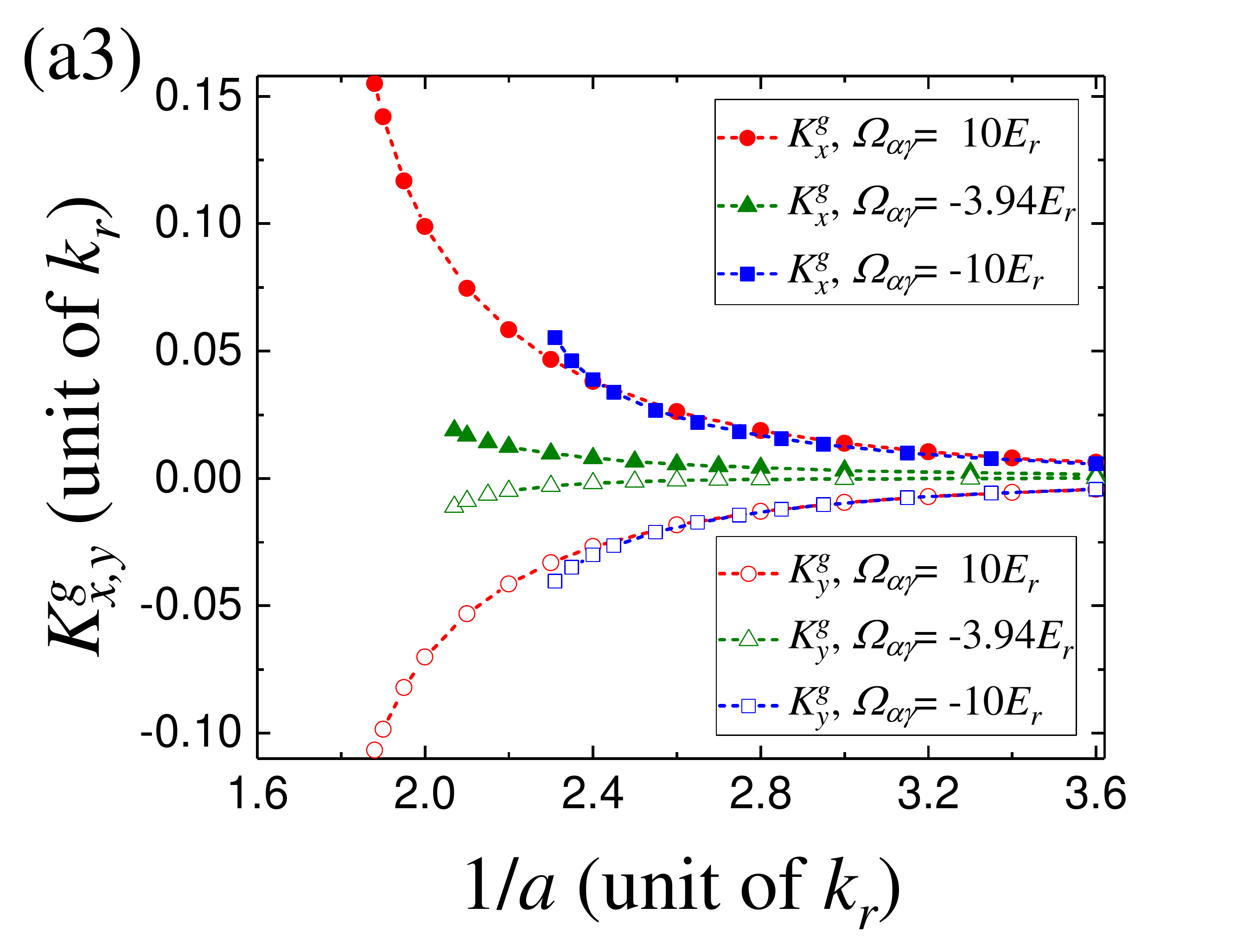}
\includegraphics[width=4cm]{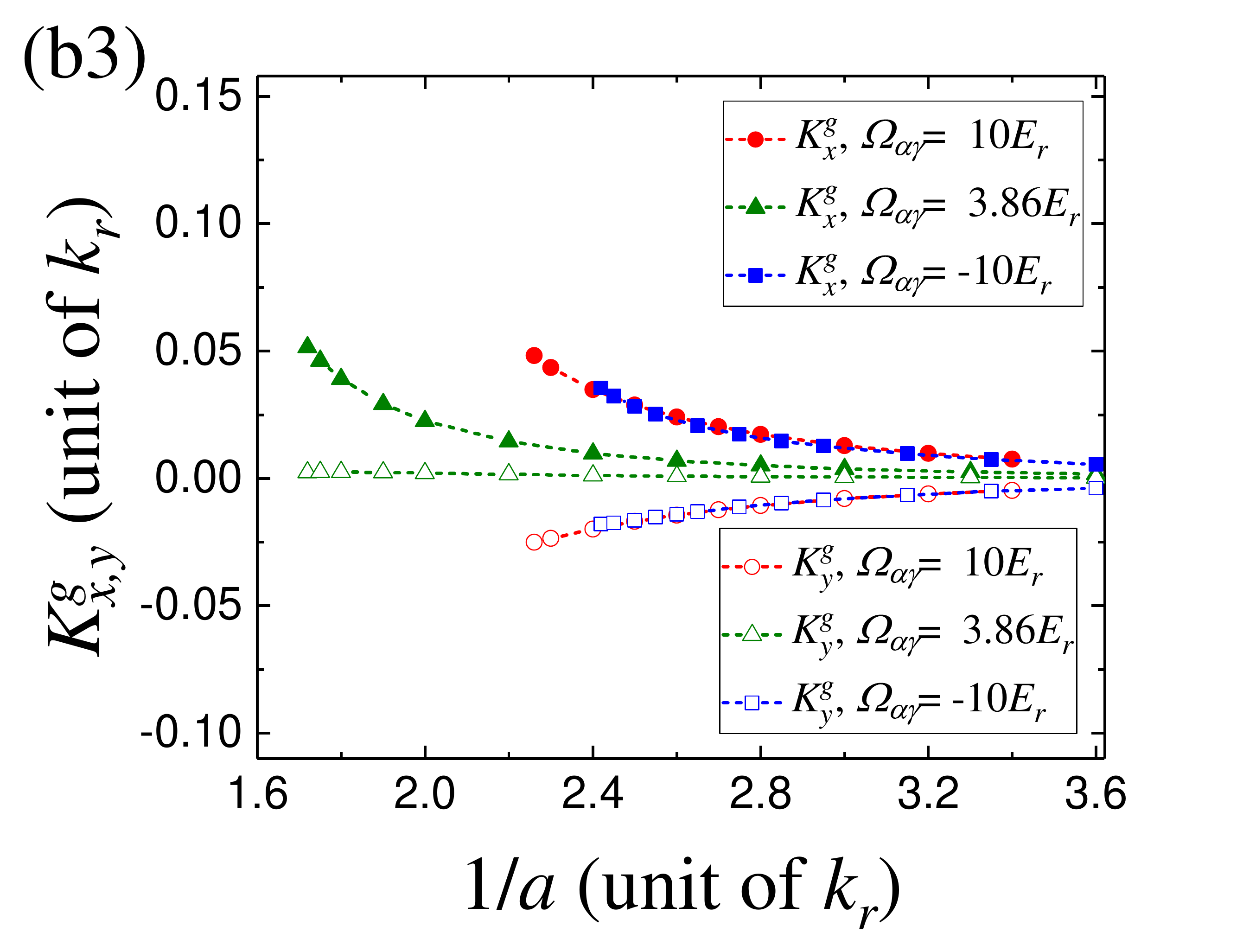}
\par\end{centering}
\caption{
(color online) {\bf (a1)-(a3):}
The properties of two-body bound state for the systems with
real effective Rabi frequencies, as in Ref.~\cite{zhangjingNP2016}.
{\bf (a1):} The thresholds $C_{{\rm th}}^{\rm min}$ and $D_{\rm th}$ as functions of $\Omega_{\alpha\gamma}$. {\bf (a2)}:
Binding energy $E_{\rm binding}^{g}$ as a function of $1/a$, for $\Omega_{\alpha\gamma}=-3.94E_r$, $\Omega_{\alpha\gamma}=-10E_r$ and $\Omega_{\alpha\gamma}=10E_r$, as well as the curve $E_{\rm binding}^{g}=-1/(ma^2)$ for the case without Raman beams (orange dashed-dotted line).
{\bf (a3)}: The CoM momentum $K^g_x$ (filled symbols) and $K^g_y$ (empty symbols) of the ground two-body bound state, as a function of $1/a$, for $\Omega_{\alpha\gamma}=10E_r$ (red circles), $\Omega_{\alpha\gamma}=-3.94E_r$  (green triangles) and $\Omega_{\alpha\gamma}=-10E_r$  (blue squares).
All the other parameters ((a1)-(a3)) are same as in Fig.~\ref{ck} (a).
{\bf (b1)-(b3):} The properties of two-body bound state for the systems with
complex effective Rabi frequencies, as in Ref.~\cite{zhangjingPRL2016}.
 {\bf (b1):} The thresholds  $C_{{\rm th}}^{\rm min}$ and $D_{\rm th}$ as functions of $\Omega_{\alpha\gamma}$. {\bf (b2)}: Binding energy
$E_{\rm binding}^{g}$ as a function of $1/a$,
for $\Omega_{\alpha\gamma}=3.86E_r$, $\Omega_{\alpha\gamma}=-10E_r$ and $\Omega_{\alpha\gamma}=10E_r$, as well as the curve $E_{\rm binding}^{g}=-1/(ma^2)$ for the case without Raman beams (orange dashed-dotted line). {\bf (b3)}:  The CoM momentum $K^g_x$ (filled symbols) and $K^g_y$ (empty symbols) of the ground two-body bound state, as a function of $1/a$, $\Omega_{\alpha\gamma}=10E_r$ (red circles), $\Omega_{\alpha\gamma}=3.86E_r$  (green triangles) and $\Omega_{\alpha\gamma}=-10E_r$  (blue squares).
All the other parameters of ((b1)-(b3)) are same as in Fig.~\ref{ckc} (a).
 \label{ebas}}
\end{figure}

\section{properties of two-body bound state}\label{sec:4}

In this section we investigate properties of the two-body bound state.
For the convenience of our discussion, we define
 the two-body bound state with the {\it lowest} energy in the ${\bf K}$-space as the {\it ground bound state}, and denote the energy of this bound state as ${\cal E}_b^{\rm ground}$. Namely, for any given scattering length $a$ and the laser parameters,  ${\cal E}_b^{\rm ground}$ is the minimum value of the bound-state energy $E_b({\bf K})$ in the ${\bf K}$-space, i.e., \begin{eqnarray}
{\cal E}_b^{\rm ground}={\rm Min}[E_b({\bf K})].
\end{eqnarray}
In addition, we introduce two parameters $C_{{\rm th}}^{\rm min}$ and $E_{{\rm th}}^{\rm min}$, which are defined as the minimum values of the threshold $C_{{\rm th}}(K_x,K_y)$ and the threshold energy $E_{{\rm th}}({\bf K})$  in the ${\bf K}$-space, respectively. That is, we have
\begin{eqnarray}
C_{{\rm th}}^{\rm min}\equiv{\rm Min}[C_{{\rm th}}(K_x,K_y)],
\end{eqnarray}
and
\begin{eqnarray}
E_{{\rm th}}^{\rm min}\equiv{\rm Min}[E_{{\rm th}}({\bf K})].
\end{eqnarray}

According to the above definitions, when the condition $1/a>C_{{\rm th}}^{\rm min}$ is satisfied, two-body bound states can appear in some region of the ${\bf K}$-space. However, as ${\cal E}_b^{\rm ground}$ and $E_{{\rm th}}^{\rm min}$ typically occur at different CoM momentum ${\bf K}$, ${\cal E}_b^{\rm ground}$ can only be lower than $E_{{\rm th}}^{\rm min}$ when $1/a$ is larger than another critical value $D_{\rm th}$, with $D_{\rm th}>C_{\rm th}^{\rm min}$, i.e., we have
\begin{eqnarray}
E_b^{\rm ground}<E_{\rm th}^{\rm min},\ \ {\rm for}\ \frac{1}{a}>D_{\rm th}.
\end{eqnarray}
In Fig.~\ref{ebas} (a1)(b1), we illustrate $C_{\rm th}^{\rm min}$ and $D_{\rm th}$ as functions of the Rabi frequencies under different experimental schemes.

Importantly, stable two-body bound states only exist for $1/a>D_{\rm th}$.
On one hand, when
\begin{align}
C_{\rm th}^{\rm min}<\frac{1}{a}<D_{\rm th},\label{c2}
\end{align}
the energies of all the two-body bound states are higher than $E_{{\rm th}}^{\rm min}$, and ${\cal E}_b({\bf K})$ crosses $E_{\rm th}({\bf K})$ in the ${\bf K}$-space.
It is clear that $E_{{\rm th}}^{\rm min}$ is the minimal eigen-energy of the two-body free Hamiltonian $H_F$, which is the lower bound of the energies of the two-body scattering states. Thus, under the condition (\ref{c2}), for any two-body bound state, there are always some scattering states with lower energies, albeit at different CoM momentum than the bound state. It follows that, in a many-body system, these two-body bound states are unstable, and can decay to the scattering-state continuum via three-atom or four-atom collisions.
On the other hand, when the condition
\begin{eqnarray}
\frac{1}{a}>D_{\rm th}\label{c3}
\end{eqnarray}
is satisfied, ${\cal E}_b^{\rm ground}$ would be lower than the minimal energy $E_{{\rm th}}^{\rm min}$ of the scattering states, and the ground bound state is stable.

In the following, we investigate the property of the ground bound state under the condition (\ref{c3}). For convenience, we denote the CoM momentum corresponding to the ground bound state as ${\bf K}^g=(K_x^g,K_y^g,K_z^g)$.
We also define the binding energy of the ground bound state as
\begin{eqnarray}
E_{\rm binding}^g\equiv E_{\rm th}({\bf K}^g)-{\cal E}_b({\bf K}^g).
 \end{eqnarray}
In the absence of SOC, we have ${\bf K}^g=0$ and $E_{\rm binding}^g=1/(ma^2)$.
For our systems with Raman-induced 2D SOC,
as shown in Fig.~\ref{ebas}(a2)(b2), $E_{\rm binding}^g$ becomes smaller than $1/(ma^2)$, i.e., the bound state becomes shallower. Furthermore, in the presence of SOC, we still have $K_z^g=0$, whereas $(K_x^g, K_y^g)$ becomes non-zero, since the ground two-body bound state has non-zero CoM momentum, as illustrated in
in Fig.~\ref{ebas}(a3)(b3).

 The existence of ground bound state with a finite CoM momentum suggests the emergence of Fulde-Ferrell pairing states in a many-body setting. Similar to SOC-induced Fulde-Ferrell states discussed previously~\cite{socreview5}, the appearance of the finite CoM momentum pairing is due to the explicit breaking of rotational symmetry in the single-particle spectrum.

At the end of this section, we emphasis that, since the threshold $C_{\rm th}({\bf K})$ tends to positive infinity in the limit ${|K_y|\rightarrow+\infty}$  or ${K_x\rightarrow+\infty}$, for any given scattering length $a$, the condition $1/a>C_{\rm th}(K_x,K_y)$ can never  be satisfied when $|K_y|$ or $K_x$ is too large, and the two-body bound state only appears for sufficiently small $|K_y|$ and $K_x$.

\section{Summary}\label{sec:5}

In summary, we have studied two-body bound states for fermions under the Raman-induced two-dimensional SOC and with $s$-wave interactions. While the presence of SOC reduces the stability of two-body bound state, the ground two-body bound state acquires a finite CoM momentum in the $x-y$ plane as SOC breaks rotational symmetry of the single-particle spectrum. Based on these results, we expect that for a many-body system on the mean-field level, competition between Fulde-Ferrell pairing states and normal state should give rise to a rich phase diagram. Such a competition is induced by the interplay of the two-dimensional SOC in the dressed-state basis and contact interaction in the hyperfine-spin basis.
In the future, it would be interesting to further explore the stability of a possible topological Fulde-Ferrell pairing state in the system.

\begin{acknowledgments}
This work is is supported in part by the National Key Research and Development Program
of China (Grant No. 2018YFA0306502 (PZ), 2016YFA0301700 (WY), 2017YFA0304100 (WY)), NSFC Grant No. 11434011 (PZ),
No. 11674393 (PZ), as well as the Research Funds of Renmin University
of China under Grant No. 16XNLQ03 (PZ).
\end{acknowledgments}

\appendix

\section{Derivation of Eq.~(\ref{equa})}

In this appendix we show how to derive Eq.~(\ref{equa}) in the main
text. Here our notations for various Hilbert spaces and the quantum
states in each space are same as the ones in Sec.~II. Substituting
Eq.~(\ref{h}) into the Schr$\ddot{{\rm o}}$dinger equation (\ref{se}),
we obtain
\begin{equation}
|\Psi_{b}\rangle\rangle=\frac{1}{{\cal E}_{b}-H_{F}}U|\Psi_{b}\rangle\rangle.\label{lsb}
\end{equation}
Furthermore, the interaction $U$ defined in Eq.~(\ref{U})
can be re-expressed as a separable form
\begin{equation}
U=U_{0}|S\rangle\langle S|\otimes|\phi)(\phi|,\label{u2}
\end{equation}
with the state $|\phi)$ being defined as
\begin{equation}
|\phi)=\frac{1}{(2\pi)^{\frac{3}{2}}}\int_{k<k_{c}}d{\bf k}|{\bf k}).\label{phi}
\end{equation}
Substituting Eq.~(\ref{u2}) into Eq.~(\ref{lsb}) we further derive
\begin{equation}
\chi_{b}=U_{0}\langle S|(\phi|\frac{1}{{\cal E}_{b}-H_{F}}|\phi)|S\rangle\chi_{b},\label{chi}
\end{equation}
with $\chi_{b}=\langle S|(\phi|\Psi_{b}\rangle\rangle$. Eq.~(\ref{chi})
yields
\begin{equation}
\frac{1}{U_{0}}=\langle S|(\phi|\frac{1}{{\cal E}_{b}-H_{F}}|\phi)|S\rangle.\label{u02}
\end{equation}

On the other hand, as shown in Sec.~II, we define the two-body internal
state $|\Lambda,{\bf k},{\bf K}\rangle\in{\cal H}_{s1}\otimes{\cal H}_{s2}$
($\Lambda=1,2,...,9$) and the energy $E_{\Lambda,{\bf k},{\bf K}}$
as the $\Lambda$-th eigen-state of the operator $h({\bf k},{\bf K})\equiv H_{{\rm 1b}}\left(\frac{{\bf K}}{2}+{\bf k}\right)+H_{{\rm 1b}}\left(\frac{{\bf K}}{2}-{\bf k}\right)$
and the corresponding eigen-energy, respectively. Thus, the eigen-state
of operator $H_{F}$, which should be a vector in the complete Hilbert
space ${\cal H}$, is the product state $|\Lambda,{\bf k},{\bf K}\rangle|{\bf k})$,
and the corresponding eigen-energy is just $E_{\Lambda,{\bf k},{\bf K}}$.
Namely, we have
\begin{equation}
H_{F}|\Lambda,{\bf k},{\bf K}\rangle|{\bf k})=E_{\Lambda,{\bf k},{\bf K}}|\Lambda,{\bf k},{\bf K}\rangle|{\bf k}).\label{hfeigen}
\end{equation}
Using this result, we can re-express the operator $1/({\cal E}_{b}-H_{F})$
as
\begin{equation}
\frac{1}{{\cal E}_{b}-H_{F}}=\int d{\bf k}\sum_{\Lambda=1}^{9}\frac{|{\bf k})({\bf k}|\otimes|\Lambda,{\bf k},{\bf K}\rangle\langle\Lambda,{\bf k},{\bf K}|}{{\cal E}_{b}-E_{\Lambda,{\bf k},{\bf K}}}.\label{green}
\end{equation}
Substituting Eq.~(\ref{green}) and the renormalization relation (\ref{renor})
into Eq.~(\ref{u02}), we obtain
\begin{equation}
\frac{1}{(2\pi)^{3}}\int_{k<k_{c}}d{\bf k}J[{\cal E}_{b},{\bf k};{\bf K}]=\frac{1}{4\pi a},\label{eq:}
\end{equation}
with the funciton $J[{\cal E}_{b},{\bf k};{\bf K}]$ being defined
in Eq.~(\ref{jjj}). Taking $k_{c}\rightarrow\infty$ for Eq.~(\ref{eq:}),
we immediately derive Eq.~(\ref{equa}).

\section{$C_{\rm th}$ under other parameters}

\begin{figure}[t]
\begin{centering}
\includegraphics[width=4cm]{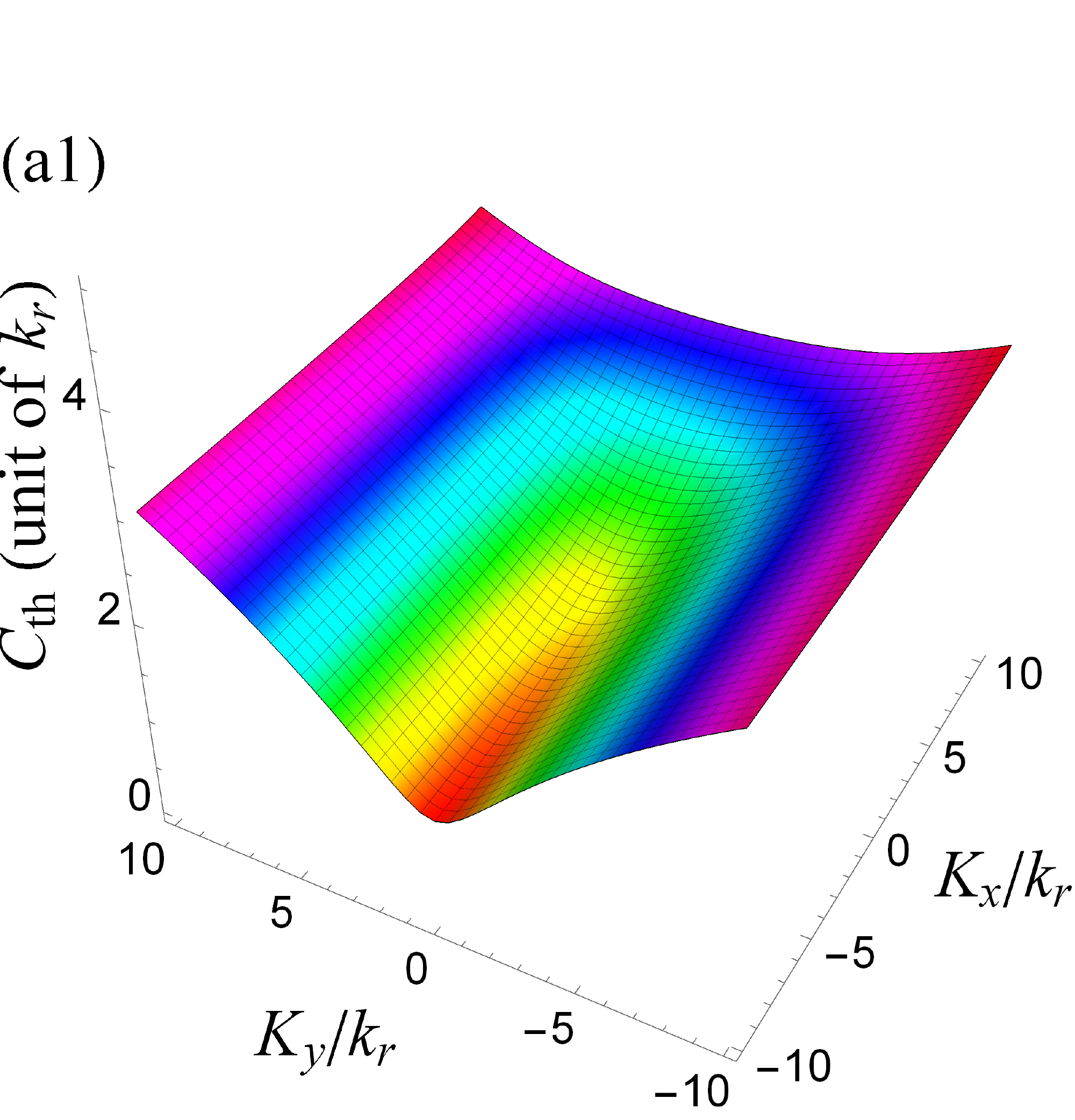}
\includegraphics[width=4cm]{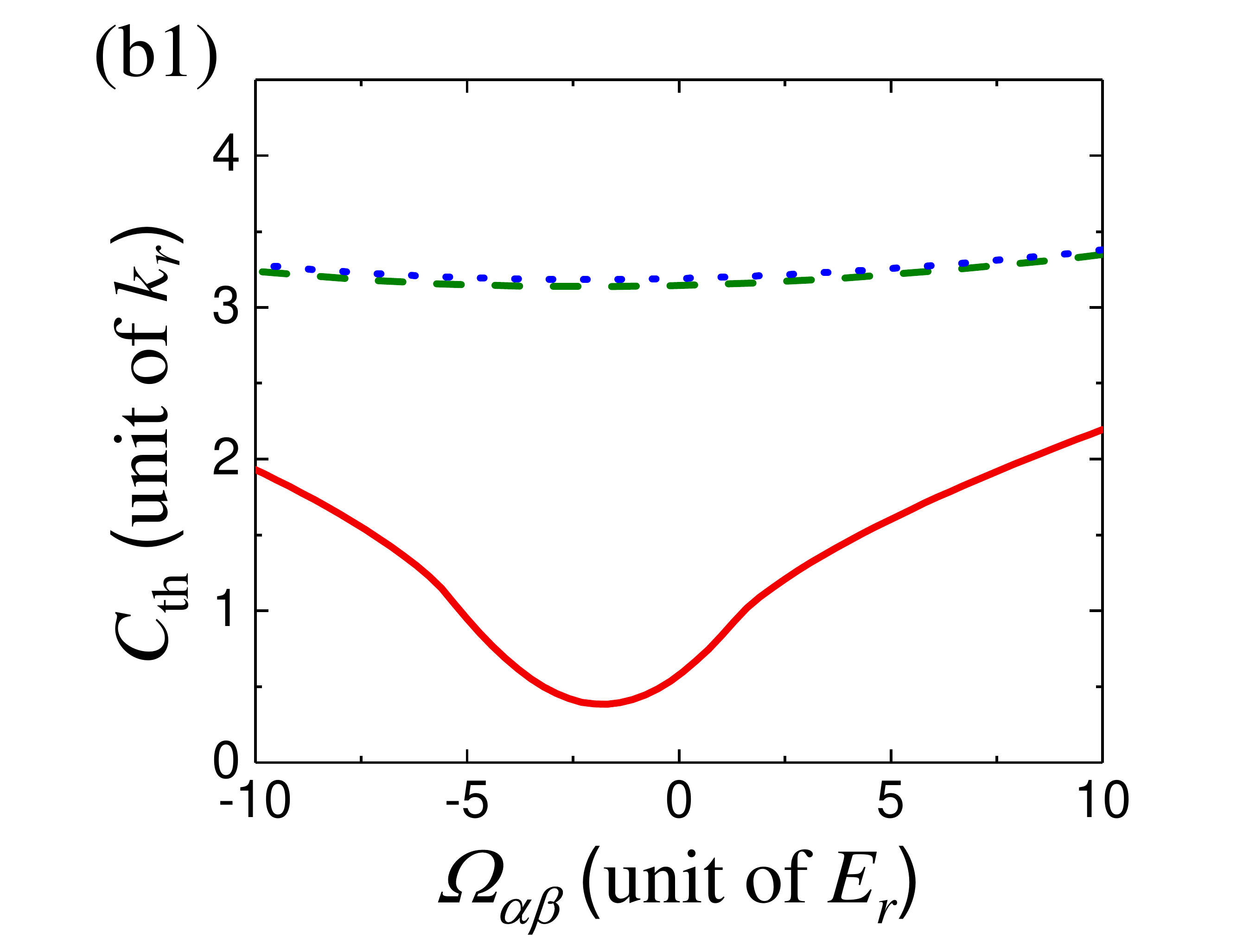}
\includegraphics[width=4cm]{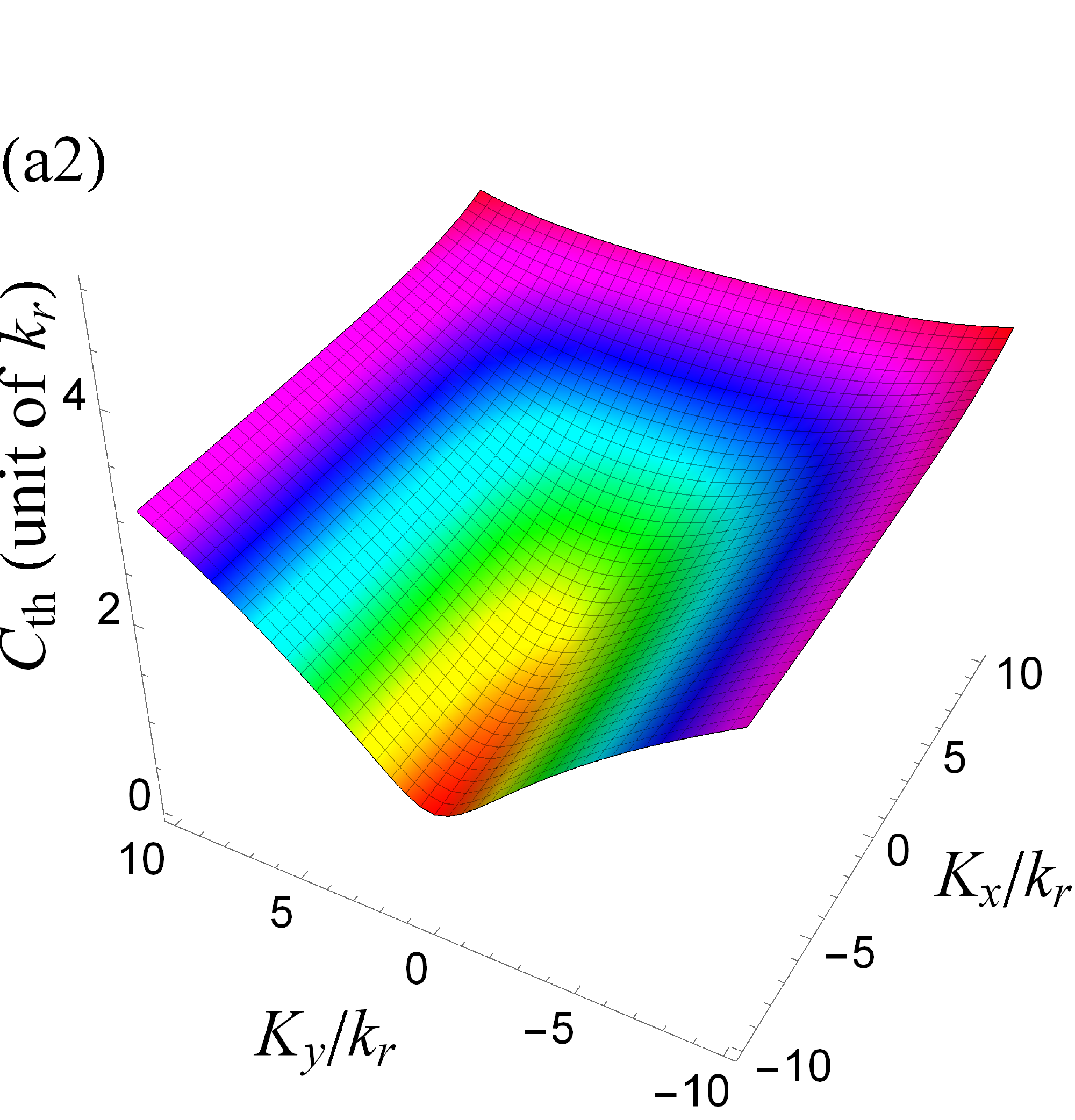}
\includegraphics[width=4cm]{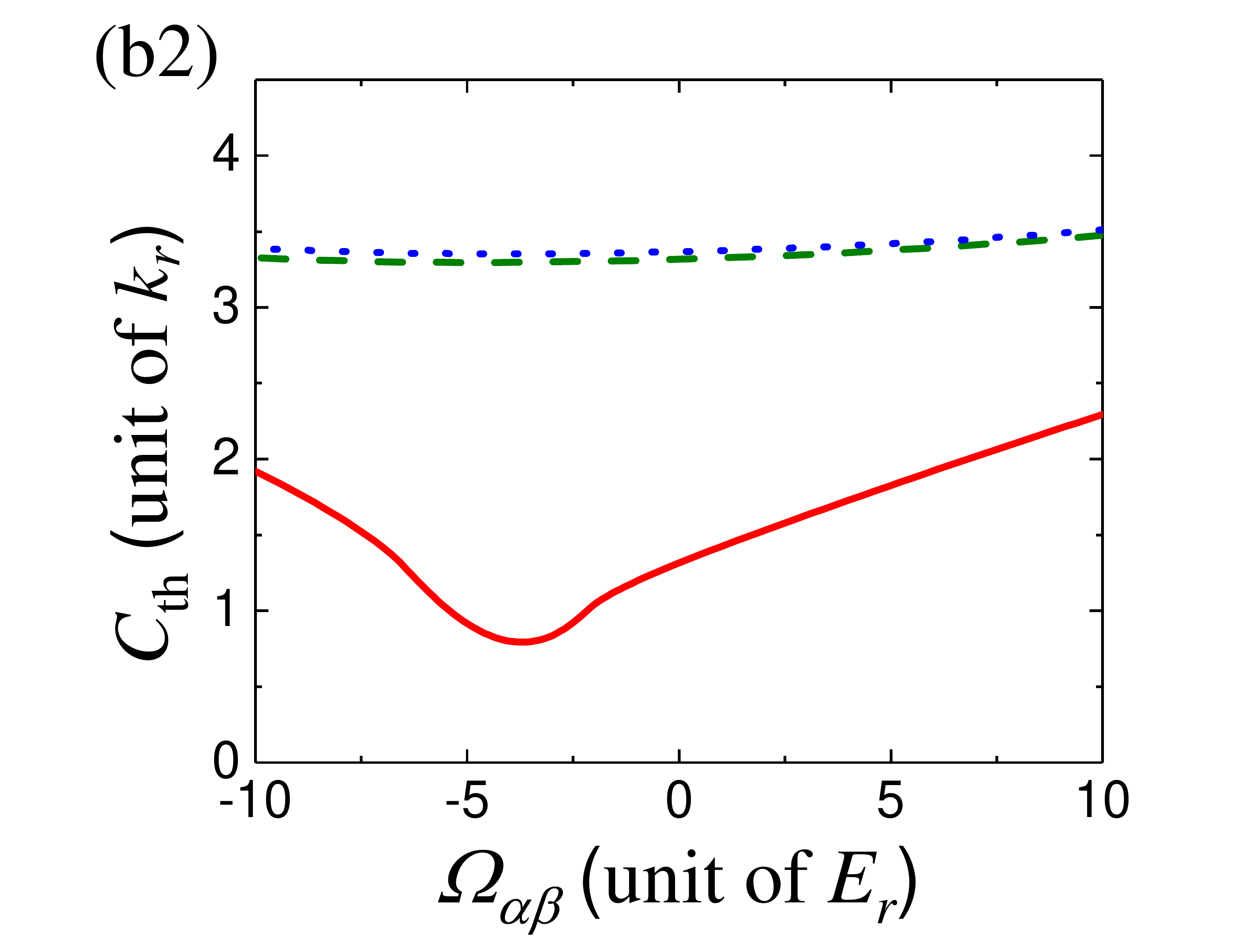}
\includegraphics[width=4cm]{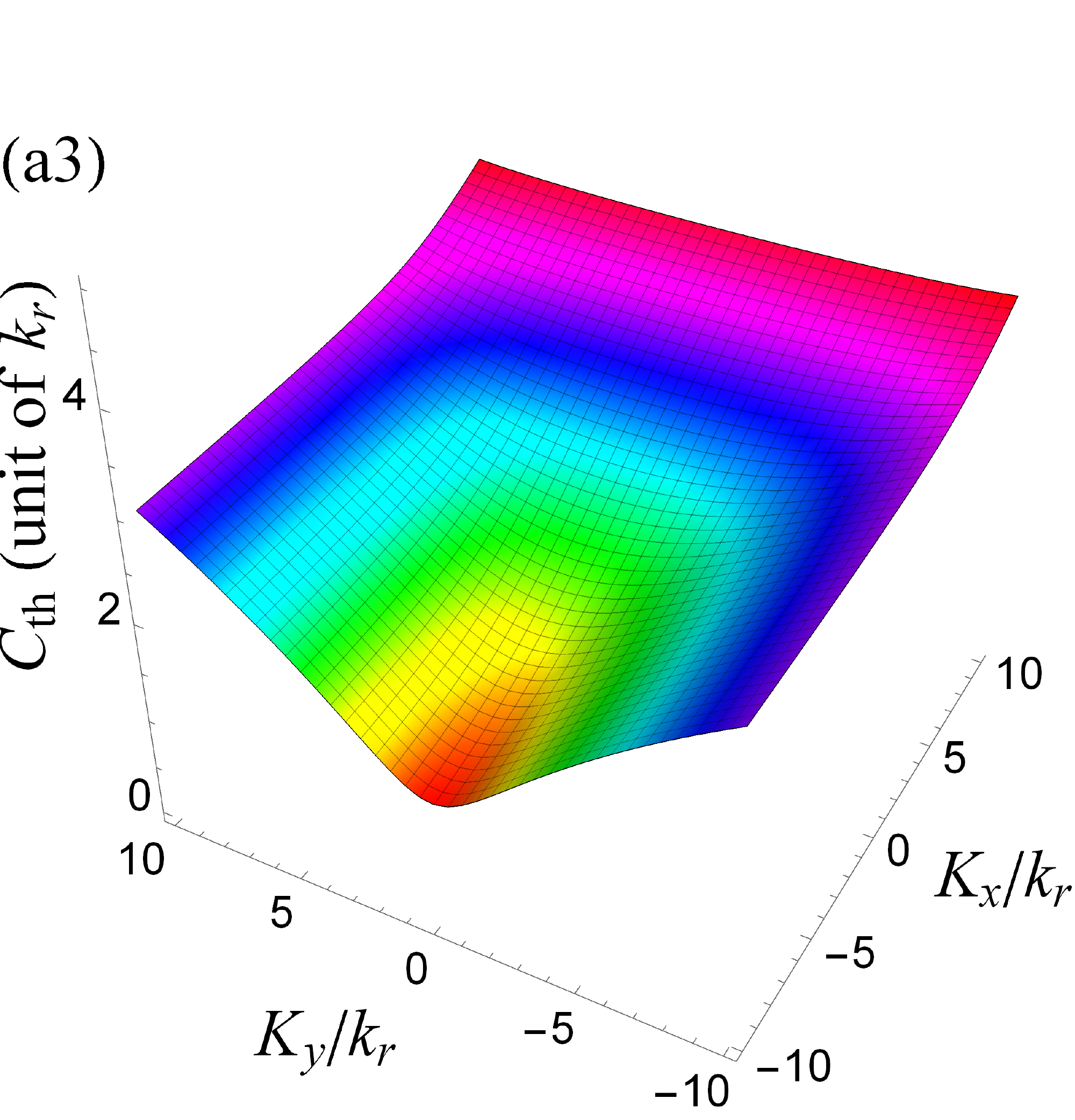}
\includegraphics[width=4cm]{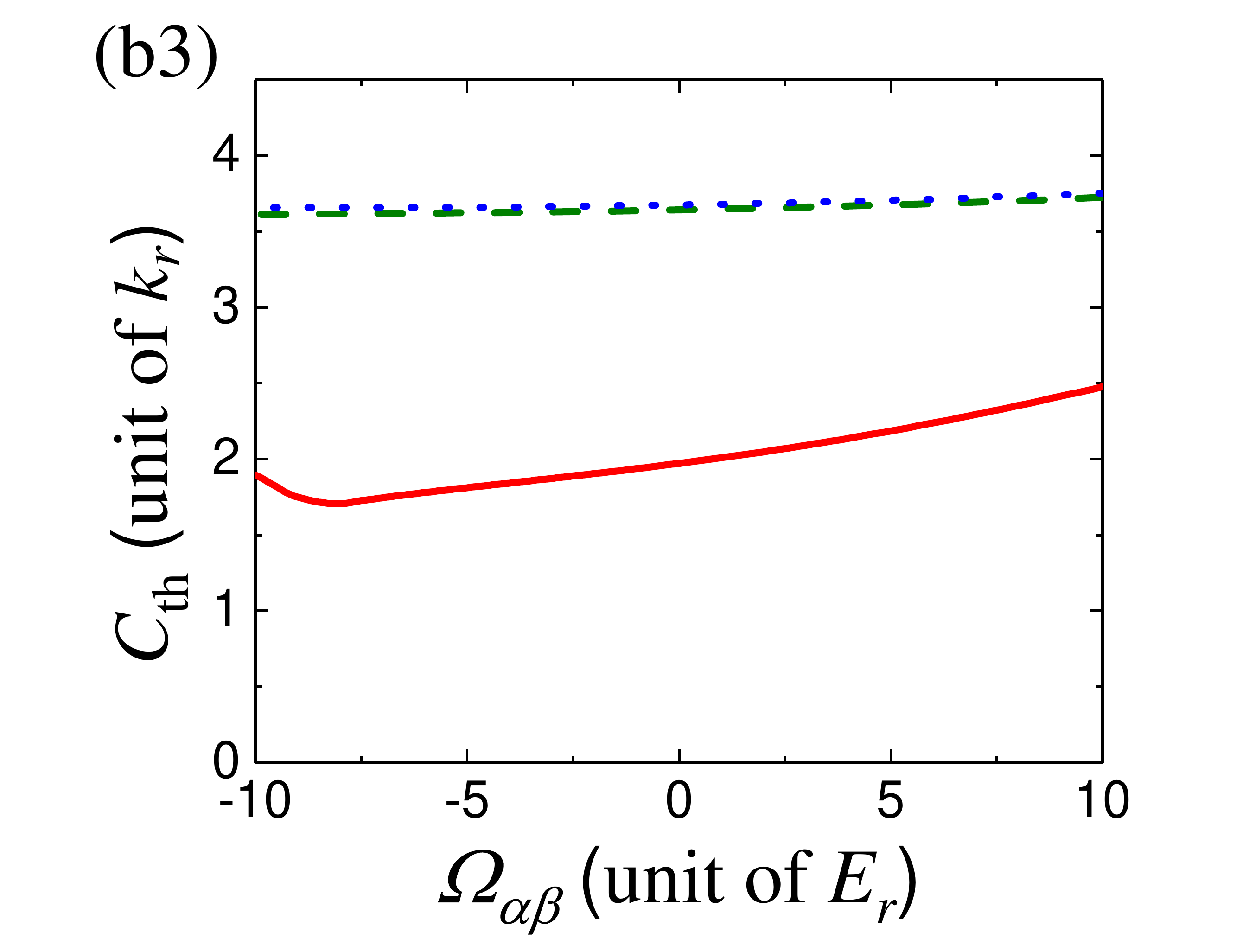}
\par\end{centering}
\caption{(color online) (\textbf{(a1-a3)}): The  two-body bound state threshold  $C_{\rm th}$
as a function of CoM momentum $(K_{x},K_{y})$,
with
real effective Rabi frequencies. For all subplots, we have
$\delta_{\beta}=0E_{r}$, $\delta_{\gamma}=3.23E_{r}$ (a1);
$\delta_{\beta}=0E_{r}$, $\delta_{\gamma}=0E_{r}$ (a2);
$\delta_{\beta}=0E_{r}$, $\delta_{\gamma}=-3.23E_{r}$ (a3), and other parameters are same as those in Fig.~\ref{ck}.
\textbf{(b1-b3)}: The threshold  $C_{\rm th}$
as a function of the effective Rabi frequency $\Omega_{\alpha\beta}$,
for cases with CoM momentum $(K_x,K_y)=(0,0)$ (red solid line), $(K_x,K_y)=(10k_r,10k_r)$ {(green dashed line)}, and $(K_x,K_y)=(10k_r,-10k_r)$ (blue dotted line). Other parameters of (b1), (b2) and (b3) are same as (a1), (a2) and (a3), respectively.
\label{appck}}
\end{figure}

\begin{figure}[t]
\begin{centering}
\includegraphics[width=4cm]{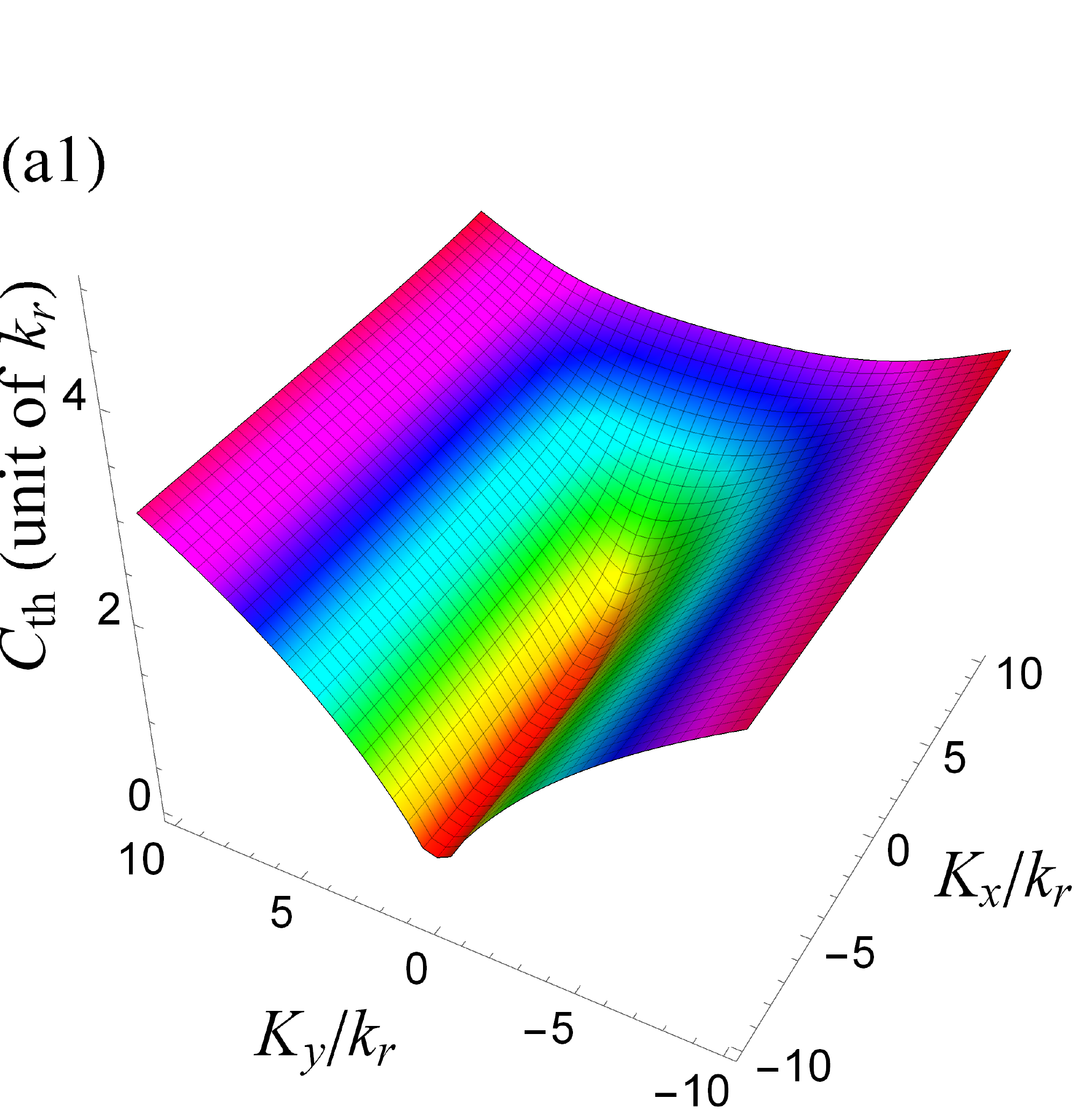}
\includegraphics[width=4cm]{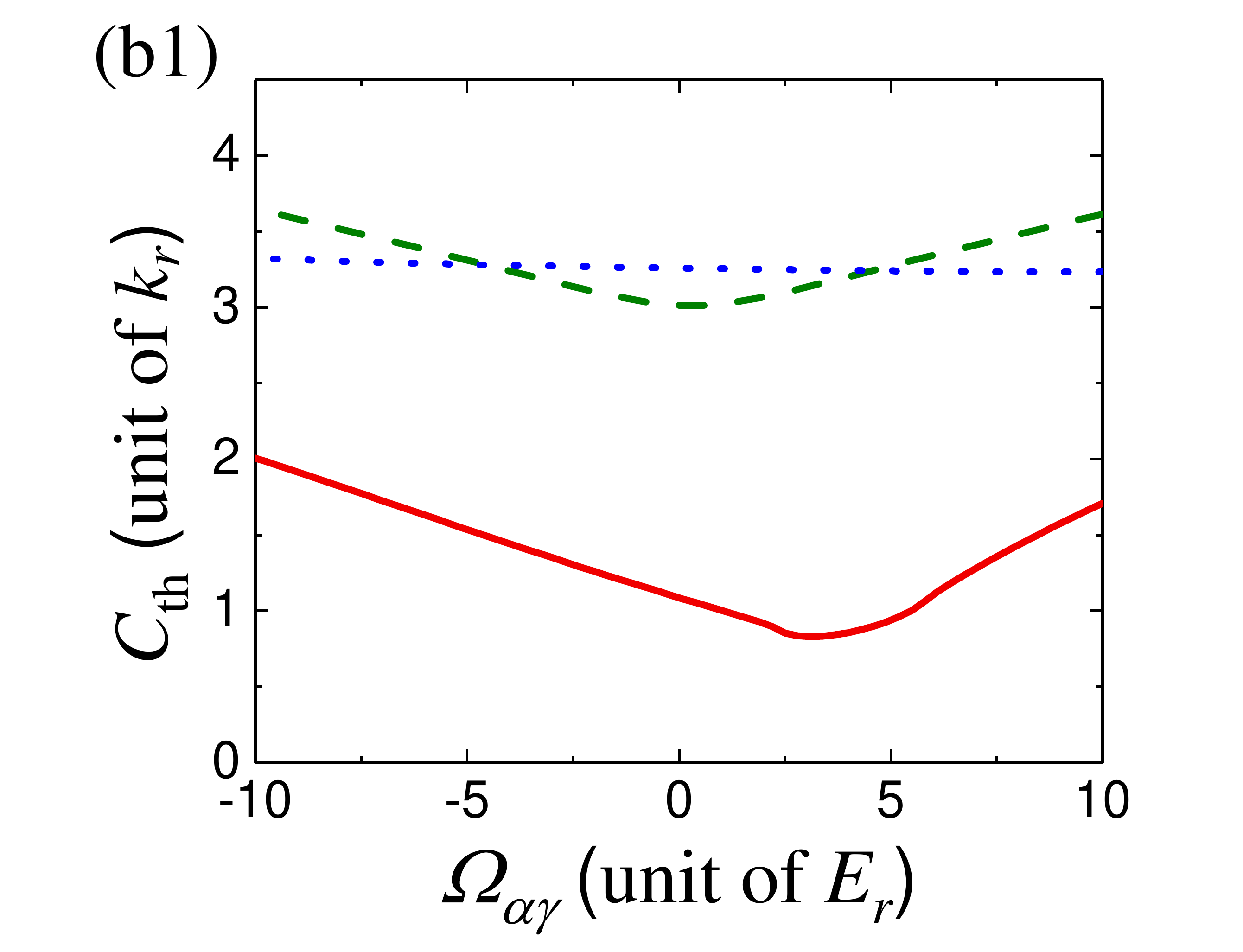}
\includegraphics[width=4cm]{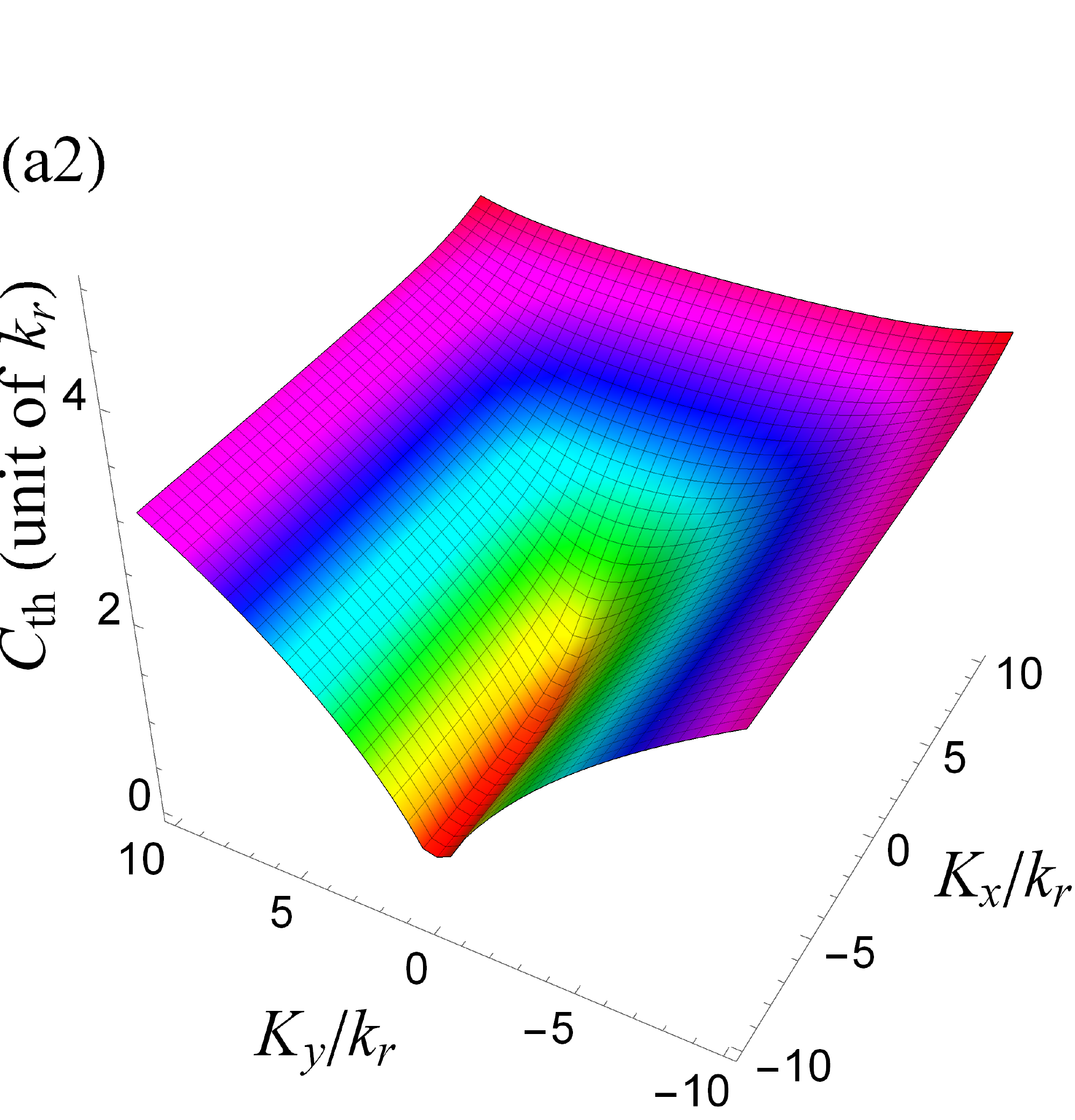}
\includegraphics[width=4cm]{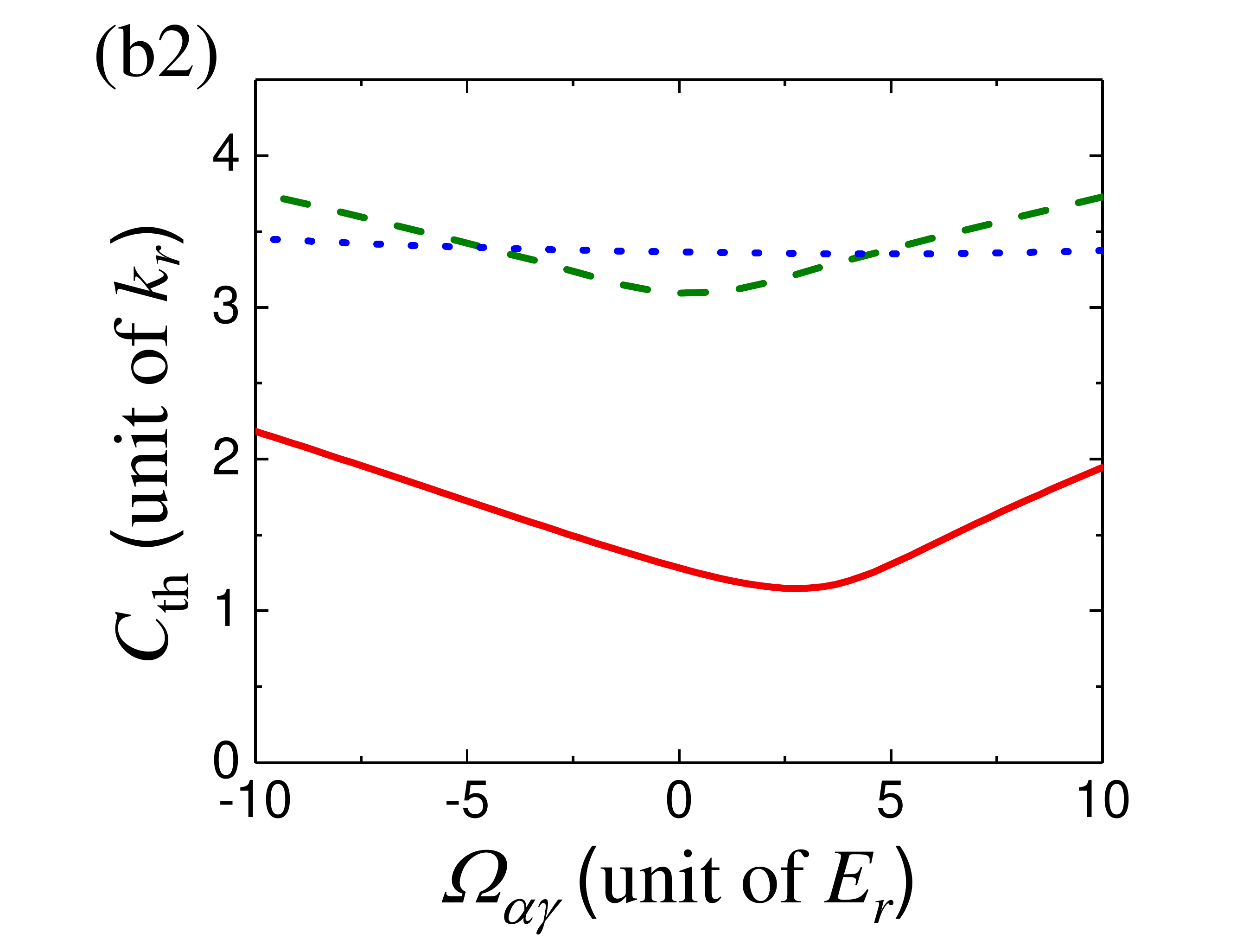}
\includegraphics[width=4cm]{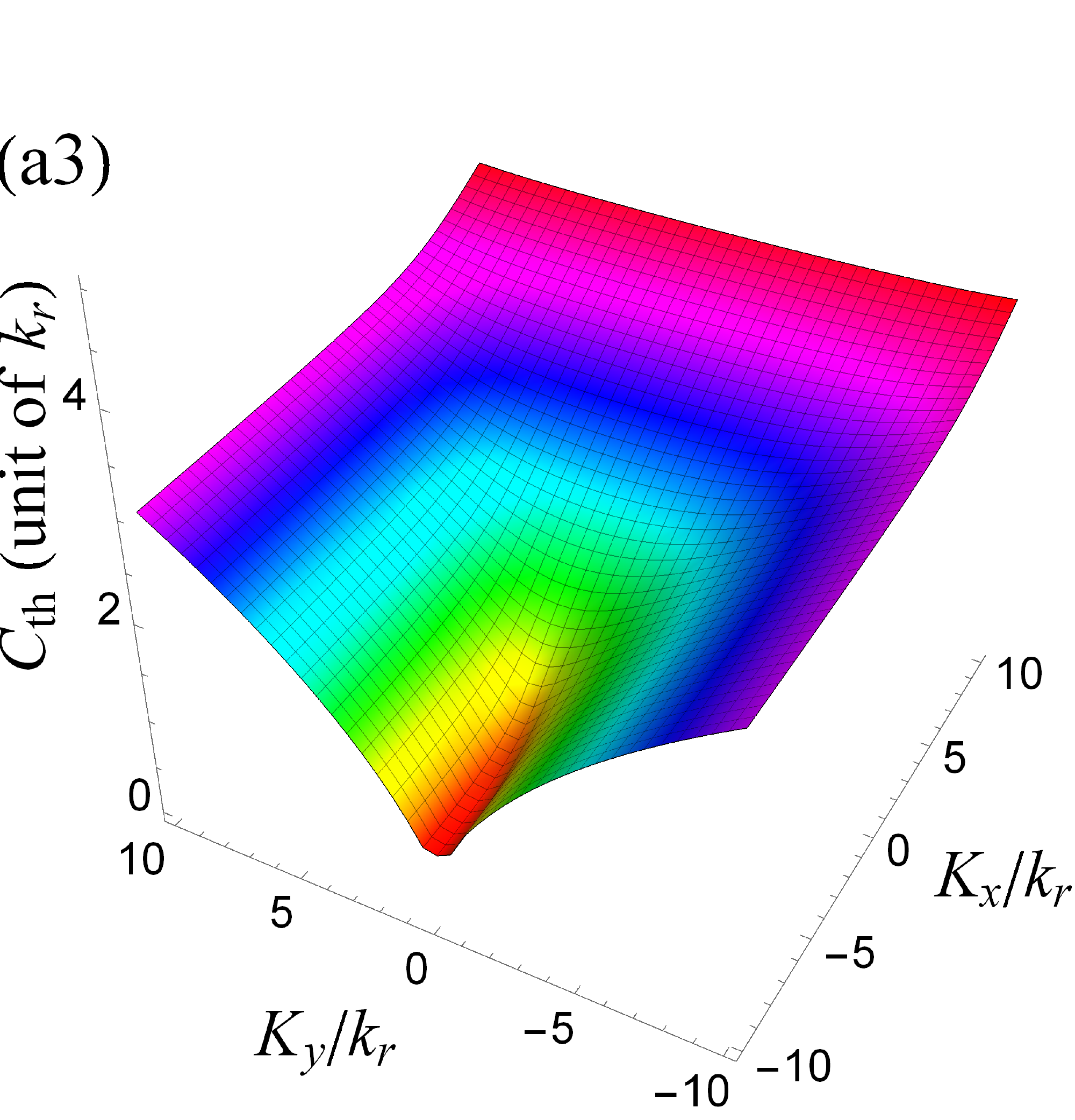}
\includegraphics[width=4cm]{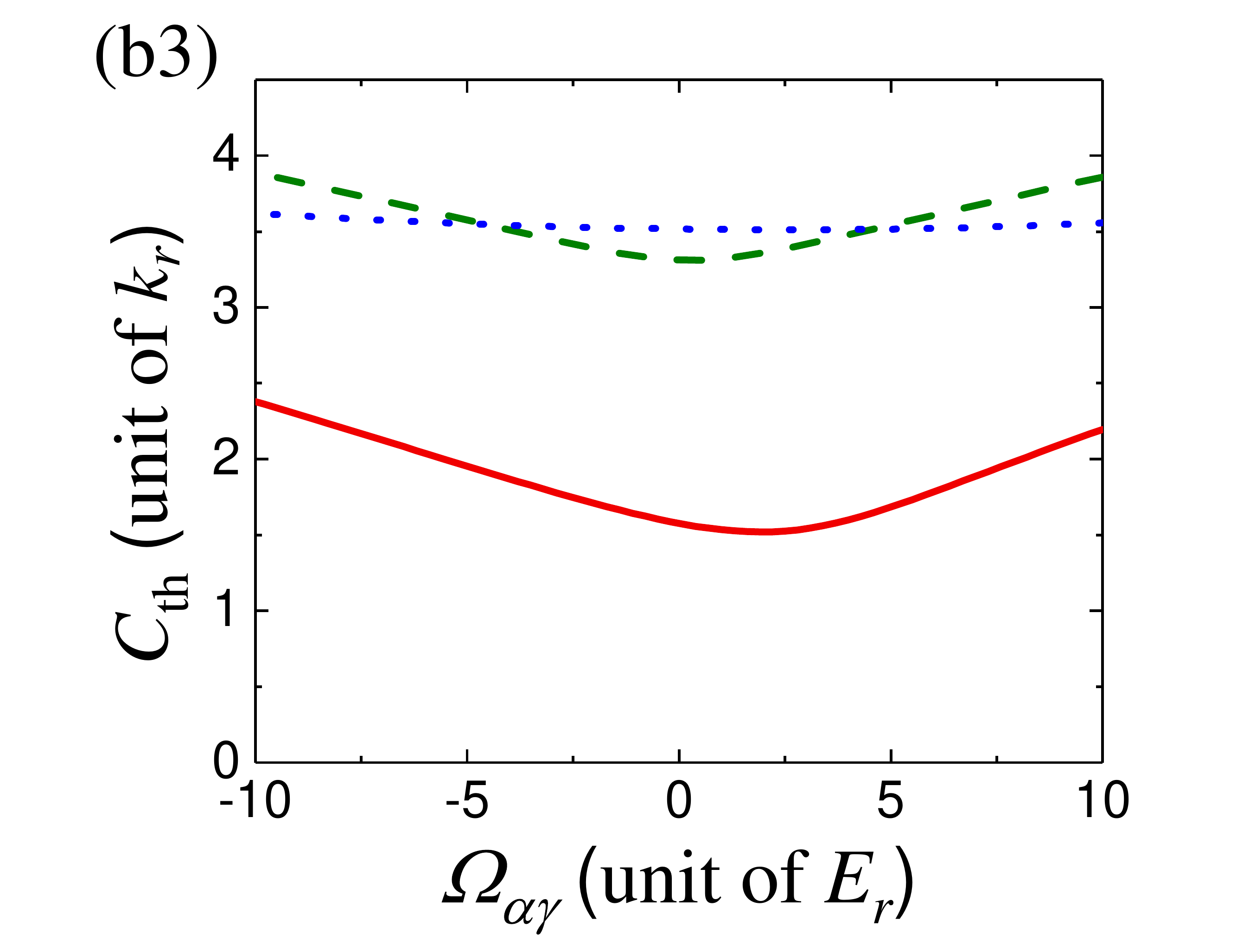}
\par\end{centering}
\caption{(color online)
(\textbf{(a1-a3)}): The  two-body bound state threshold  $C_{\rm th}$
as a function of CoM momentum $(K_{x},K_{y})$,
with
complex effective Rabi frequencies.
For all subplots, we have
$\delta_{\beta}=0E_{r}$, $\delta_{\gamma}=1.8E_{r}$ (a1);
$\delta_{\beta}=0E_{r}$, $\delta_{\gamma}=0E_{r}$ (a2);
$\delta_{\beta}=0E_{r}$, $\delta_{\gamma}=-1.8E_{r}$ (a3), and other parameters are same as those in Fig.~\ref{ckc}.
\textbf{(b1-b3)}: The threshold $C_{\rm th}$
as a function of the effective Rabi frequency $\Omega_{\alpha\gamma}$,
for cases with CoM momentum $(K_x,K_y)=(0,0)$ (red solid line), $(K_x,K_y)=(10k_r,10k_r)$ { (green dashed line)}, and $(K_x,K_y)=(10k_r,-10k_r)$ (blue dotted line). Other parameters of (b1), (b2) and (b3) are same as (a1), (a2) and (a3), respectively.
\label{appckc}}
\end{figure}

As mentioned in Sec.~III, we numerically calculate the threshold $C_{\rm th}$ of the two-body bound state, following experimental systems with real~\cite{zhangjingNP2016} and complex~\cite{zhangjingPRL2016} effective Rabi frequencies. Some results are shown in Fig.~\ref{ck} and Fig.~\ref{ckc} of the main text. Here we illustrate
the results for more cases with various Rabi frequencies $\Omega_{\alpha\beta(\alpha\gamma)}$ and detunings $\delta_{\beta(\gamma)}$ in Fig.~\ref{appck} and Fig.~\ref{appckc}. All our results support our conclusion in the main text that $C_{\rm th}$ is shifted to positive side of the $1/a$-axis.

\section{Threshold of bound state for large $K_{y}$ and $K_{x}$}

As illustrated in Sec.~III, the threshold $C_{\rm th}$ of the two-body
bound state becomes very large in the limit $|K_{y}|\rightarrow\infty$ or
$K_{x}\rightarrow+\infty$.
This is because, in such
limit, the threshold energy $E_{{\rm th}}$ becomes very low,
which
makes the bound state difficult to form. In this appendix, we
show some detailed analysis supporting this picture.


For the convenience of our discussion,
here we denote the Hilbert space where the atoms are in the two-body internal states $|i\rangle_1|j\rangle_2$ ($i,j=\alpha,\beta$) as ${\cal H}_{\alpha\beta}$, and denote the space with the atoms being in $|\gamma\rangle_1|j\rangle_2$ or  $|j\rangle_1|\gamma\rangle_2$ ($j=\alpha,\beta$) as ${\cal H}_{\gamma}$. It is clear that the total Hilbert space ${\cal H}$
is the direct sum of these two spaces, i.e., ${\cal H}={\cal H}_{\alpha\beta}\oplus{\cal H}_{\gamma}$.

Furthermore, we can re-express the total Hamiltonian $H$ as
\begin{eqnarray}
H & = & H_{1}+H_{2}+H_{3}
\end{eqnarray}
where
\begin{eqnarray}
H_{2} & = & -\sum_{\xi,\eta=\alpha,\beta,\gamma}\frac{\Omega_{\xi\eta}}{2}\left(|\xi\rangle_{1}\langle\eta|+|\xi\rangle_{2}\langle\eta|\right),\\
H_{3} & = & -\sum_{\xi=\alpha,\beta,\gamma}\left(\frac{{\bf p}\cdot{\bf k}_{\xi}}{m}\right)\left(|\xi\rangle_{1}\langle\xi|-|\xi\rangle_{2}\langle\xi|\right),
\end{eqnarray}
and
\begin{equation}
H_{1}\equiv H-H_{2}-H_{3}.\label{hhhh1}
\end{equation}

We first consider the Hamiltonian
$H_{1}$. It is clear that
 $H_{1}$ does not include the coupling
between the states in ${\cal H}_{\alpha\beta}$ nor the states in ${\cal H}_{\gamma}$. Thus, we have
\begin{equation}
H_{1}=H_{\alpha\beta}+H_{\gamma},\label{hh1-1}
\end{equation}
with $H_{\alpha\beta}$ and $H_{\beta}$ being the operators of the
spaces ${\cal H}_{\alpha\beta}$ and ${\cal H}_{\gamma}$, respectively,
and can be expressed as
\begin{eqnarray}
H_{\alpha\beta} & = & \sum_{\xi,\eta=\alpha,\beta}\left[\frac{{\bf p}^{2}}{m}+\tilde{\delta}_{\xi}+\tilde{\delta}_{\eta}\right]\otimes|\xi\rangle_{1}\langle\xi|\otimes|\eta\rangle_{2}\langle\eta|+U\nonumber\\
&\equiv& H_{\alpha\beta}^{(F)}+U;\label{hab}\\
H_{\gamma} &=&  \sum_{\eta=\alpha,\beta}\left[\frac{{\bf p}^{2}}{m}+\tilde{\delta}_{\gamma}+\tilde{\delta}_{\eta}\right]\otimes\nonumber\\
&&\left(|\gamma\rangle_{1}\langle\gamma|\otimes|\eta\rangle_{2}\langle\eta|+|\eta\rangle_{1}\langle\eta|\otimes|\gamma\rangle_{2}\langle\gamma|\right),
\label{hg}
\end{eqnarray}
where
\begin{equation}
\tilde{\delta}_{\alpha}=\delta_{\alpha}-\frac{K_{y}k_{r}}{2m};\ \tilde{\delta}_{\beta}=\delta_{\beta}+\frac{K_{y}k_{r}}{2m};\ \tilde{\delta}_{\gamma}=\delta_{\gamma}-\frac{K_{x}k_{r}}{2m}.\label{dt}
\end{equation}

\begin{figure}[t]
\begin{centering}
\includegraphics[width=9cm]{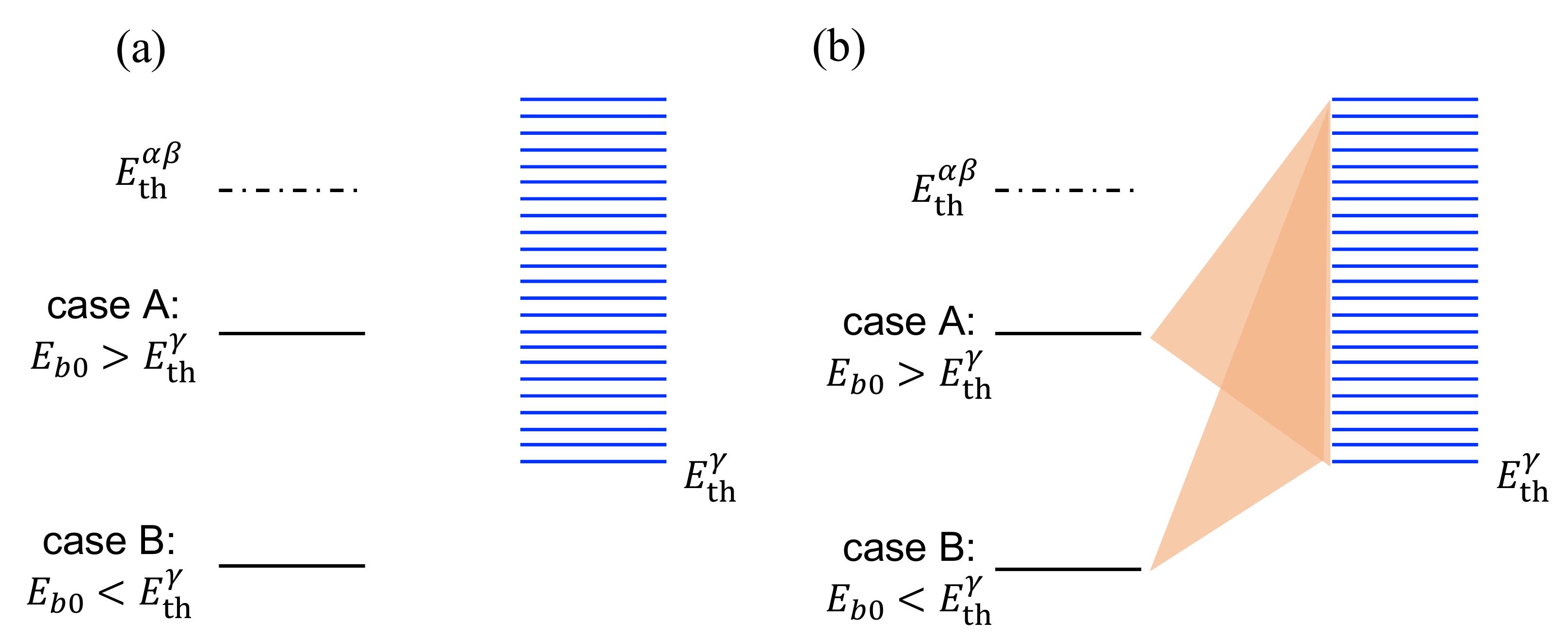}
\par\end{centering}
\caption{(color online)  {\bf (a)}: Some eigen-levels of $H_1$. The black solid line indicates the bound-state contributed by $H_{\alpha\beta}$ while the blue solid line indicates continuous spectrum contributed by $H_{\gamma}$.
In addition, $E_{\rm th}^{\alpha\beta}$ (black dashed-dotted line) is the lower bound of other continuous eigen-levels contributed by the projection of $H_{\alpha\beta}$ in the subspace  where the two atoms are in different internal states~\cite{explain}.
If $H_2$ and $H_3$ were zero, the bound state would always be stable, whether the energy $E_{b0}$ be higher (case A) or lower (case B) than the lower bound $E_{\rm th}^{\gamma}$ of the continuous spectrum of $H_{\gamma}$, since the bound state is not coupled to states in the continuum. {\bf (b)}: In the presence of $H_2$, the bound state is coupled to the continuous spectrum of $H_{\gamma}$, with the coupling being shown as brown shadow. As a result, the bound state is stable only if the energy is lower than the  lower bound $E_{\rm th}^{\alpha\beta}$ of this spectrum.\label{appfig}}
\end{figure}

Therefore, if both $H_{2}$ and $H_{3}$ were zero, the two atoms
could form a bound state only if $1/a>0$, which is supported by $H_{\alpha\beta}$
and can be denoted by $|\Phi_{b0}\rangle\rangle$. The energy of this
bound state is
\begin{equation}
E_{b0}=E_{{\rm th}}^{\alpha\beta}-\frac{1}{ma^{2}},\label{eb0}
\end{equation}
where
\begin{equation}
E_{{\rm th}}^{\alpha\beta}=\delta_{\alpha}+\delta_{\beta}
\end{equation}
is the minimum eigen-energy of the projection of $H_{\alpha\beta}^{(F)}$ (i.e., the
``free Hamiltonian in the space ${\cal H}_{\alpha\beta}$'')
in the subspace  where the two atoms are in different internal states~\cite{explain}.
On
the other hand, as shown in Fig.~\ref{appfig}(a), the Hamiltonian $H=H_{1}$ also have many
eigen-states
with continuous eigen-energies, which are eigen-states of $H_{\gamma}$.
The lower-bound of this continuous spectrum is
\begin{equation}
E_{{\rm th}}^{\gamma}\equiv{\rm Min}[\tilde{\delta}_{\gamma}+\tilde{\delta}_{\beta},\tilde{\delta}_{\gamma}+\tilde{\delta}_{\alpha}].
\end{equation}
 Here we emphasis that, the bound-state energy $E_{b0}$ may be either
higher or lower than $E_{{\rm th}}^{\gamma}$. Nevertheless,
the bound state $|\Phi_{b0}\rangle\rangle$ is always stable because
it is not coupled to these continuous states $H_{\gamma}$.

Now we consider the effect from $H_{2}$. For simplicity, here
we also ignore $H_{3}$. As shown in
Fig.~\ref{appfig}(b),
the Hamiltonian $H_{2}$ can induce the coupling
between the bound state $|\Phi_{b0}\rangle\rangle$ and the continuous
eigen-levels of $H_{\gamma}$. In this case, the bound state is no longer
stable when energy $E_{b0}$ is higher than the lower-bound $E_{{\rm th}}^{\gamma}$
of this continuous spectrum, while it is still stable when $E_{b0}$
is lower than $E_{{\rm th}}^{\gamma}$. Therefore, in the presence
of $H_{2}$, the two atoms can form a bound state only when $E_{b0}<E_{{\rm th}}^{\gamma}$
or
\begin{eqnarray}
\frac1{ma^{2}}>E_{{\rm th}}^{\alpha\beta}-E_{{\rm th}}^{\gamma}.\label{con2}
\end{eqnarray}

Furthermore, when $|K_{y}|$ is very large, $E_{{\rm th}}^{\gamma}$ would be much lower than
$E_{{\rm th}}^{\alpha\beta}$.
That is due to the term $W$ defined in Eq.~(\ref{ww}), which is now a part of $H_\gamma$.
As a result, only the very deep bound state could be stable. Explicitly, in this case
the condition (\ref{con2}) can be approximately expressed as
\begin{eqnarray}
\frac 1a>\sqrt{k_{r}\frac{|K_{y}|}{2}},
\end{eqnarray}
and thus
the threshold $C_{\rm th}$ is approximately $\sqrt{k_{r}{|K_{y}|}/{2}}$, which increases with $|K_{y}|$, as shown
in Sec. III.

Similarly, when $K_{x}$ takes a {\it positive} large value, due to the term $-{K_{x}k_{r}}/{(2m)}$ in the expression (\ref{dt}) of $\tilde{\delta}_{\gamma}$, $E_{{\rm th}}^{\gamma}$ would also be much lower than
$E_{{\rm th}}^{\alpha\beta}$. As a result, the threshold  $C_{\rm th}$ can be approximately expressed as $C_{\rm th}\approx\sqrt{k_{r}{K_{x}}/{2}}$, and increases with $K_x$. On the other hand, when $K_{x}$ takes a {\it negative} large value, $E_{{\rm th}}^{\gamma}$ becomes much higher than
$E_{{\rm th}}^{\alpha\beta}$. Namely, the energy of the space ${\cal H}_\gamma$ is much higher than the bound state of $H_{\alpha\beta}$. Thus, the threshold  $C_{\rm th}$ of the two-body bound state is hardly affected by the inter-space coupling $H_2$ but tends to a constant, as shown in  Fig.~\ref{ck}(a, c) and Fig.~\ref{ckc}(a, c).

Finally, we consider the effect of $H_{3}$. Since $H_{3}$ is independent
of ${\bf K}$ and does not include the coupling between the spaces
${\cal H}_{\alpha\beta}$ and ${\cal H}_{\gamma}$, and only induces
a ${\bf K}$-independent modification for the threshold energies $E_{{\rm th}}^{\alpha\beta}$
and $E_{{\rm th}}^{\gamma}$, as well as the expressions of the bound
state $E_{b0}$. Therefore, in the presence of $H_{3}$, our above
analysis is still qualitatively correct.



\begin{thebibliography}{1}
\bibitem{uu} Here we also emphasize that in the limit $k_{c}\rightarrow\infty$
we have $\frac{1}{(2\pi)^{3}}\int_{k,k'<k_{c}}|{\bf k})({\bf k}'|d{\bf k}d{\bf k}'=\delta({\bf r})$,
where ${\bf r}$ is the relative-position operator of the two atoms,
and thus $U=U_{0}\delta({\bf r})|S\rangle\langle S|$. This is why
such an effective interaction is called ``renormalized contact
interaction''.
\bibitem{explain} For our system, the Hamiltonian is close in the subspace where the two atoms are in different internal states, and the bound-state is also an element of this subspace. Thus, only the projection of $H_{\alpha\beta}$ in this subspace is relevant for our problem.
\end{thebibliography}
\end{document}